\documentclass[fleqn,usenatbib]{mnras}

\usepackage{newtxtext,newtxmath}
\usepackage[T1]{fontenc}

\DeclareRobustCommand{\VAN}[3]{#2}
\let\VANthebibliography\thebibliography
\def\thebibliography{\DeclareRobustCommand{\VAN}[3]{##3}\VANthebibliography}

\usepackage{graphicx}	% Including figure files
\usepackage{amsmath}	% Advanced maths commands
\usepackage{ae,aecompl}
\usepackage{mathrsfs}
\usepackage{color}
\usepackage{enumerate}
\usepackage{booktabs}
\usepackage{soul}
\usepackage{array}
\usepackage{mathalfa} 
\usepackage{subcaption}

\usepackage{float} % in order to use [H]
\usepackage{dcolumn}% Align table columns on decimal point
\usepackage{bm}% bold math
\usepackage{graphicx}
\title[Reliability of Quasars as Cosmological Probes]{On the Reliability of Quasars as Cosmological Distance Indicators}

\author[Ariadna Montiel et al.]{
Ariadna Montiel,$^{1}$
Sofia Samario-Nava,$^{2}$\thanks{E-mail: ssamario@icf.unam.mx}
Juan Carlos Hidalgo$^{2}$
and Jose Ignacio Cabrera$^{3}$ $^{4}$
\\
$^{1}$Physics Department, Centro de Investigaci\'{o}n y de Estudios Avanzados del Instituto Polit\'ecnico Nacional (Cinvestav), PO. Box 14-740, Mexico City, Mexico\\
$^{2}$Instituto de Ciencias F\'isicas, Universidad Nacional
Aut\'onoma de M\'exico,  62210, Cuernavaca, Morelos, Mexico\\
$^{3}$Facultad de Ciencias, Universidad Nacional Aut\'onoma de M\'exico,  04510, Ciudad de M\'exico, Mexico\\
$^{4}$Colegio de Ciencias y Humanidades Plantel Sur, Universidad Nacional Autónoma de México, 04500 Ciudad de México, México}

\date{Accepted XXX. Received YYY; in original form ZZZ}

\pubyear{\the\year{}}

\begin{document}
\label{firstpage}
\pagerange{\pageref{firstpage}--\pageref{lastpage}}
\maketitle

\begin{abstract}

We assess the viability of quasars as cosmological distance indicators based on the non-linear $L_X$--$L_{\rm UV}$ relation. We calibrate this relation in a model-independent way by anchoring quasar luminosity distances to cosmic-chronometer $H(z)$ measurements over $z\leq 1.43$, and construct a quasar Hubble diagram extending up to $z\simeq 7.5$. We compare a traditional stepwise approach, in which the calibration is fixed before cosmological inference, with a joint QSO+SN calibration--cosmology framework where calibration and cosmological parameters are sampled simultaneously. The stepwise analysis, supplemented with DESI DR2 BAO measurements and Planck compressed CMB distance priors, is used as a diagnostic benchmark, while the joint framework, with and without the SH0ES $H_0$ information, provides our main cosmological results.

We selected a low-$z$ quasar subsample ($z<1.43$), matching the redshift range of the cosmic-chronometer calibration, and found calibration parameters consistent with previous studies. However, the stepwise cosmological constraints can become unstable once quasars are included, reflecting the incomplete propagation of calibration uncertainties and calibration--cosmology degeneracies. In contrast, the joint analysis yields self-consistent constraints because the quasar calibration parameters are fitted simultaneously with the cosmological parameters, allowing these uncertainties and degeneracies to be propagated into the final posteriors. Our results indicate that the current limitations of quasar cosmology are driven mainly by intrinsic scatter and possible sample-dependent effects in the $L_X$--$L_{\rm UV}$ relation, rather than by a fundamental inconsistency with standard cosmology. 

\end{abstract}

\begin{keywords}
quasars -- standard candles -- dark energy -- 
\end{keywords}

%%%%%%%%%%%%%%%%%%%%%%%%%%%%%%%%%%%%%%%%%%%%%%%%%%

%%%%%%%%%%%%%%%%% BODY OF PAPER %%%%%%%%%%%%%%%%%%

\section{Introduction}

Type Ia supernovae (SNe Ia) are fundamental to the standard cosmological model, allowing us to trace the expansion of the Universe over the past 10 billion years \citep{Riess_2007}. They provided the first direct evidence for the accelerating expansion of the Universe \citep{Riess:1998cb} and continue to be essential tools for studying cosmological models in a range of redshifts $z < z^{\mathrm{SNeIa}}_{\mathrm{max}}\sim 2.3$ \citep{Scolnic:2017caz,Scolnic_2022}. To investigate cosmic evolution at higher redshifts, observations of distant and luminous objects are required. This high redshift regime is only weakly sampled by current probes, and cosmological models remain poorly tested in this range.   

Quasars are among the most promising candidates for cosmological distance indicators at high redshifts, since observations range up to $z \sim 7$ \citep{Mortlock_2011}. They are extremely variable anisotropic sources characterized by a wide range of luminosities. Unlike SNe Ia \citep{Hachinger2008}, there is no direct connection between any specific spectral observable and their luminosity. Therefore, in order to employ these objects in cosmology, a way to measure a “standard luminosity” is needed to estimate their distances. 

Over the years, various strategies have been proposed to exploit quasars as \textit{standarizable candles} by leveraging their diverse emission properties. An early attempt was made by \cite{Baldwin1977}, who identified a correlation between the widths of quasar emission lines and their luminosities, suggesting a potential standardization method. However, the viability of this approach was later questioned by \cite{Osmer:1998mb}, who highlighted the significant scatter in the correlation, limiting its usefulness for cosmological purposes. A different avenue was explored in \cite{PhysRevLett.110.081301}, where the luminosity behavior of super-Eddington accreting black holes was analyzed. The authors found that their emission is predominantly determined by black hole mass, proposing these systems as potential standard candles. Similarly, \cite{Franca_2014} showed that the X-ray variability of quasars can be used to measure their luminosities, providing another possible route to distance estimation. While multiple approaches have been proposed to standardize quasars, the correlation between the logarithmic X-ray and UV luminosities \citep{Tananbaum1979,Zamorani1981,Lusso2010,Vagnetti2010} has gained particular attention to measure cosmological distances, as discussed by \cite{Risaliti_2015,Risaliti:2016nqt,Lusso2017AA,Risaliti:2018reu,Lusso2020F,Khadka:2020whe}.

In \cite{Lusso_2016} quasars appeared to be in good agreement with the $\Lambda$CDM model when calibrated using the non-linear relation between their ultraviolet and X-ray luminosities, although this calibration was model dependent. Subsequent studies revealed tensions at higher redshifts; for example, \cite{Risaliti:2018reu} found that the best-fit for matter density $\Omega_m$ differs when using quasars at low and high redshifts. Nevertheless, \cite{Banerjee:2020bjq} demonstrated that this discrepancy arises because a cosmographic expansion had been employed beyond its range of validity.

 Also, \cite{Khadka2021} showed that the calibration parameters $\gamma$ and $\beta$ depend on the assumed cosmology, concluding that quasars are reliable distance indicators only up to $z \lesssim 1.5-1.7$. In a subsequent paper, \cite{Khadka:2021xcc} analyzed these quasar data
in greater detail to determine whether specific sub-samples are
mainly responsible for the redshift- and model-dependent behavior of the $\gamma$ and $\beta$ values. They found that the largest of seven sub-samples exhibits $L_X-L_{UV}$ relations that depend on the cosmological model assumed and also on redshift, which appears to be the primary cause of the similar behavior observed in the full quasar compilation.

Additionally, \cite{Li2021} performed a model-independent calibration up to $z<2$, also finding deviations from $\Lambda$CDM for $z>2$.  However, \cite{Sacchi2022} confirmed that the UV–X-ray relation still holds for quasars beyond $z = 2.5$ using high-quality spectroscopy. Moreover, \cite{Li2022} attributed the observed tension not to new physics, but rather to redshift evolution and non-universal intrinsic dispersion in the luminosity relation. \cite{Wang_2022} constructed a three-dimensional and redshift-evolutionary X-ray and UV luminosity relation for quasars using the powerful statistical tool \texttt{copula} \cite{Nelsen2006}, showing that it yields better consistency with the data. In a follow-up study,  \cite{Wang:2024nsi} compared three forms of the  $L_X - L_{UV}$ relation and found that versions including redshift evolution continue to best match the data. 

Given its significant tension with the standard cosmological model, the quasar dataset has been extensively scrutinized and critically examined \citep{Khadka:2020vlh, Khadka:2022aeg,Zajacek:2023qjm}. Overall, these studies suggest that the tension between quasars and the $\Lambda$CDM model likely originates from an evolution in the luminosity relation. This discrepancy can be alleviated by introducing a redshift dependence in the parameters of the quasars' calibration. However, \cite{Lusso:2025bhy} challenged this interpretation, emphasizing the absence of any significant redshift evolution in the slope of the luminosity relation. Moreover, they show that the derived distance estimates are consistent with the standard flat $\Lambda$CDM model up to $z \sim 1.5$, while notable deviations appear only at higher redshifts. According to their analysis, all reported inconsistencies can be naturally attributed to limitations inherent in the cosmological model used for data interpretation.

In this work, we estimate quasar distances following the method of \cite{Risaliti_2015,Lusso_2016} within a model-independent framework. Our primary goal is to assess whether quasars can serve as reliable cosmological distance indicators at redshifts well beyond the reach of Type Ia supernovae, thereby extending the cosmic distance ladder to $z>2$. We use the 2038-object quasar sample compiled by \cite{Lusso:2020pdb} and construct the corresponding Hubble diagram up to $z\simeq 7.5$. 

The non-linear $L_X$--$L_{\rm UV}$ relation is calibrated without assuming any fiducial cosmological model. Instead, we adopt the model-independent technique introduced in \cite{Montiel_2020}, employing Cosmic Chronometer measurements to reconstruct the expansion history $H(z)$ through a Bézier polynomial up to $z=1.43$. This reconstruction provides an external distance scale that anchors the quasar luminosity calibration while avoiding circularity in the cosmological inference.

Motivated by the recent analysis of \cite{Lusso:2025bhy}, which finds no compelling evidence for significant redshift evolution in the $L_X$--$L_{\rm UV}$ relation within current uncertainties, we adopt a first-order approach in which the calibration parameters are assumed to be redshift-independent. This choice allows us to explicitly test whether a simple, non-evolving calibration is sufficient to yield consistent cosmological constraints.

We then evaluate the cosmological performance of the calibrated quasar sample by constraining $\Lambda$CDM, $w$CDM, and CPL models. We begin with a traditional stepwise strategy, in which the $L_X$–$L_{\rm UV}$ relation is calibrated first, and the resulting distances are subsequently used for cosmological inference. While this approach is useful as a diagnostic benchmark and facilitates comparison with previous studies, it can underestimate the impact of calibration uncertainties and calibration-cosmology degeneracies on the inferred cosmological parameters.

For this reason, we also implement a joint calibration–cosmology framework in which calibration and cosmological parameters are sampled simultaneously within a single likelihood. A joint treatment is particularly important for quasars because the parameters of the $L_X$–$L_{\rm UV}$ relation can be partially degenerate with the cosmological parameters that determine $d_L(z)$. In the joint framework, this covariance is explicitly captured in the posterior, so that cosmological constraints consistently reflect both measurement noise and calibration uncertainty. This setup also enables controlled tests of external information, such as the impact of including (or removing) the SH0ES-based $H_0$ constraint within the same likelihood.

The paper is organized as follows. In Section~\ref{Sec II: Calibration} we describe the quasar sample and the model-independent calibration procedure. Section~\ref{Sec:III} introduces the dark-energy models considered in this work. In Section~\ref{Sec:IV} we summarize the observational datasets and computational tools, and specify the priors and sampling strategies adopted in both the stepwise and joint analyses. Section~\ref{Sec:V} presents the calibration results and the corresponding cosmological constraints, comparing the stepwise and joint approaches and assessing the impact of quasar redshift coverage (full sample versus low-$z$ subsample). We conclude in Section~\ref{sec:conclusions}.

%%%%%%%%%%%%%%%%%%%%%%%%%%%%%%%%%%%%%%%%%%%%%%%%%%%%%%%%%%%%%%%%%%%%%%%%%%%%%%%%

\section{The quasar sample and calibration method}
\label{Sec II: Calibration}

\subsection{\label{Sample}Sample}

We use the quasar sample compiled by \cite{Lusso:2020pdb}, which includes 2421 sources spanning the redshift range $0.009 \leq z \leq7.541$, with UV rest-frame monochromatic fluxes at 2500 \textup{~\AA} and X-ray rest-frame monochromatic fluxes at 2keV. In accordance with the selection criteria proposed in \cite{Lusso:2020pdb}, we restrict the sample to quasars at $z>0.7$, ensuring that the UV fluxes are based on direct photometric data without the need for spectral extrapolation. An exception was made for 15 low-redshift sources whose 2500 \textup{~\AA} fluxes were obtained from direct photometry. Our final dataset thus consists of 2038 quasars: 2023 at $z>0.7$, and 15 AGN at very low redshift from the International Ultraviolet Explorer (IUE) in the Mikulski Archive for Space Telescopes (MAST) (see Section 2.7 in \cite{Lusso:2020pdb}).

For our analysis, we use both the full calibrated quasar dataset and a low-redshift subsample defined by sources with $z\leq 1.43$. This selection ensures consistency with the data used to perform our cosmological model independent calibration and allows for direct comparison with previous studies, such as \cite{Lusso:2025bhy}, that found this redshift range to be reliable for cosmological purposes. Additionally, we bin the sample to evaluate its behavior in the Hubble diagram; however, this binning is used solely for illustrative purposes and not in any cosmological fitting or parameter estimation. The bins are constructed with logarithmic width and defined such that each contains a sufficient number of sources to ensure statistically meaningful results.

\subsection{Calibration}
\label{subsection:calibration}
The non-linear relation between X-ray and UV luminosities is usually described as 
\begin{equation}
    \log(L_X) = \gamma \log(L_{UV}) + \beta,
    \label{Eq:nonlinear}
\end{equation}
where $\gamma$ and $\beta$ are the two quasar calibration parameters to be fitted to the data.  

Since luminosity, $L$, and flux, $F$, are related by $L = 4\pi d_L^2 F$, the non-linear relation between luminosities can be expressed in terms of fluxes as follows
\begin{equation} 
\log(F_X) = \gamma \log(F_{UV}) + (2\gamma - 2)\log(d_L) + (\gamma - 1)\log(4\pi) + \beta, 
\label{Eq:fluxdL}
\end{equation} 
where $F_{X}$ and $F_{UV}$ are the fluxes measured at fixed rest-frame wavelengths, and $d_L$ is the luminosity distance. Note that from this last equation, it is explicitly the dependence on the cosmological model through $d_L$, which is defined as
\begin{equation}
    d_{L}(\Omega_K,z)= \frac{c}{H_0} \frac{(1+z)}{\sqrt{|\Omega_K|}} \mathrm{sinn} \left[ \sqrt{|\Omega_K |} \int_{0}^{z} \frac{H_0 dz'}{H(z')} \right],
    \label{Eq:d_L}
\end{equation}
where $\Omega_K$ is the present curvature density defined as $\Omega_K\equiv -K/H_0^2 a^2 $. The symbol $\mathrm{sinn}(x)$ is defined piecewise: it takes the form $\sinh{(x)}$ when $\Omega_K>0$, $\sin{(x)}$ when $\Omega_K<0$ and reduces to $x$ in the case of a flat geometry ($\Omega_K=0$). 

In the pioneering work of \cite{Risaliti_2015}, a $\Lambda$CDM cosmology was directly assumed in Eq.~\ref{Eq:d_L} to derive the best-fitting values of $\gamma$ and $\beta$ for the non-linear relation in Eq.~(\ref{Eq:nonlinear}). In contrast, our goal is to determine the calibration without imposing any specific cosmological model at the calibration stage. To this end, we follow the procedure introduced by \cite{Montiel_2020}, originally developed for a high-quality GRB dataset, using an external reconstruction of $H(z)$ from Cosmic Chronometers (CC) with a Bézier polynomial to estimate the luminosity distance in a model-independent manner over the CC redshift range. This approach enables us to derive the distance modulus, $\mu(z)$, using Eq.~(\ref{Eq:fluxdL}) and to construct the quasar Hubble diagram without assuming a fiducial background cosmology:
\begin{equation} 
\mu(z) = \frac{5}{2(\gamma - 1)}\left[\log(F_X) - \gamma \log(F_{UV}) - \beta - (\gamma - 1)\log(4\pi(1+z)) \right]. 
\end{equation}

The resulting calibrated relation is used only in the stepwise cosmological inference; in the joint calibration--cosmology analysis, Eq.~(2) is fitted directly within the full likelihood, as described in Section~\ref{Sec:V}. The stepwise construction has an important limitation: the expansion history used to calibrate the quasars is fixed by the CC-based Bézier reconstruction, whereas in the subsequent cosmological fits the expansion history is determined by the specific model under consideration, such as $\Lambda$CDM, $w$CDM, or CPL. Thus, the method does not guarantee that the $H(z)$ used in the calibration stage coincides with the $H(z)$ preferred in the cosmological-inference stage. For this reason, the stepwise results should be interpreted as a diagnostic comparison rather than as our main self-consistent cosmological constraints.

The first step of our stepwise strategy consists of constructing a B\'ezier parametric curve of degree $n$ given by
\begin{align}
        &H_{n}(z)=\sum_{d=0}^{n} \beta_{d}h_{n}^{d}(z), &h_{n}^{d}\equiv \frac{n!(z/z_{m})^{d}}{d!(n-d)!}\left(1-\frac{z}{z_{m}} \right)^{n-d},
    	\label{Ec:Bezier}
    \end{align}
where $\beta_d$ are the coefficients of a linear combination of Bernstein basis polynomials $h_{n}^{d}(z)$, which are positive over the interval $0 \leq z/z_m \leq 1$, where $z_m$ denotes the maximum redshift in the dataset. We use Hubble parameter data from \cite{Capozziello:2017nbu}, obtained via the CC method \citep{Jimenez_2002,10.1093/mnrasl/slv037}. Accordingly, the Bezier curve constructed is a polynomial of degree $n=2$. As with any finite-order reconstruction, this choice provides a smooth interpolant to the CC measurements of $H(z)$ over the covered redshift range. This approach yields a monotonically increasing function, allowing us to identify the Hubble constant, $H_0$ with the coefficient $\beta_0$ by setting $d=0$ and $z=0$.

 Cosmic Chronometers provide a cosmology-independent method to measure the Hubble parameter, $H(z)$, by analyzing the differential age evolution of massive and passive early-type galaxies. This technique enables a direct reconstruction of the expansion history of the Universe without relying on a specific cosmological model. However, this method is not free from systematic uncertainties, primarily associated with the choice of stellar population synthesis (SPS) models, stellar metallicity, star formation history (SFH), and residual young stellar populations \citep{Moresco:2020fbm}. In our previous work, following the analysis of \cite{Moresco:2020fbm}, we accounted for these systematics by adding, in quadrature, the maximum bias reported to the Hubble parameter uncertainties in the dataset from \cite{Capozziello:2017nbu} and we carried out a detailed analysis of the impact of including or excluding these additional systematic uncertainties. We found that, to ensure robustness of the analysis, it is preferable to incorporate the maximum systematic contribution; see  \cite{Montiel_2020} for further details. Here we adopt the same approach.

For consistency with the analysis performed in \cite{Moresco:2020fbm}, we restrict our CC Hubble data to $z \leq 1.43$. Accordingly, we use a sub-sample of 28 out of the 31 $H(z)$ measurements from  \cite{Capozziello:2017nbu}. Then, by employing the sub-sample of 28 measurements of the Hubble parameter, we performed a non-linear least-squares minimization by using the Python software package \textsc{lmfit} \citep{newville_matthew_2014_11813}. The best-fit parameters obtained for the B\'ezier fit with $n=2$ are
\begin{equation}
    H_{2}(z)=\beta_{0} h_{2}^{0}(z)+\beta_{1} h_{2}^{1}(z)+\beta_{2} h_{2}^{2}(z),
    \label{Ec:Bezier-n2}
\end{equation}  
where,
\begin{eqnarray}
\beta_0 = H_0=70.81, \qquad \beta_1=81.99, \qquad \beta_2=179.02,
\label{betas:2cases}
\end{eqnarray}
and the corresponding covariance matrix is, 
\[
\mathbf{cov}=
  \begin{bmatrix}
12.99 & -20.07 & 8.88\\
-20.07 & 50.18 &  -29.92\\
8.88 & -29.926 &  53.76\\
  \end{bmatrix}.
\]

\noindent 
Note the associated uncertainties for the parameters in \eqref{betas:2cases} are encoded in the diagonal elements of the matrix.

The next step of the calibration involves using the function $H_2(z)$ to compute the luminosity, $L$, via the relation $L=4\pi d_L^2 (1+z) F$ for both the X-ray and UV bands, in order to establish the non-linear relation given by Eq. (\ref{Eq:nonlinear}). The factor $(1+z)$ is included to account for cosmological redshift effects, ensuring consistency between the observed flux $F$ and the intrinsic luminosity $L$, defined in the rest frame of the source. Furthermore, we make use of the luminosity distance $d_{L}(z)$ in a flat cosmology  defined as 
\begin{equation}
    d^{\mathrm{cal}}_{L}(z)\equiv d_{L}(z)=c(1+z)\int_{0}^{z} \frac{dz'}{H_2(z')},
    \label{Ec:dLcal}
\end{equation}
where we have added the label $\mathrm{cal}$, denoting \textit{calibrated}, to emphasize that the luminosity distance used here results from a calibration procedure based on CC Hubble data at $z<1.43$. Throughout this analysis, we assume spatial flatness ($\Omega_K=0$), in line with the latest constraints from the Planck mission, which report $\Omega_K= 0.001 \pm 0.002$ \cite{Aghanim:2018eyx}.

The final stage before constructing a Hubble diagram for our quasar sample is to set the best-fit values of the parameters $\gamma$~and~$\beta$. As an initial step, we analyzed both the full quasar sample and its binned version to test the robustness of the parameter estimation. For this purpose, we employed the \textsc{BCES} algorithm as our model-fitting technique. \textsc{BCES}, which stands for bivariate correlated errors and intrinsic scatter, was introduced by \cite{Akritas:1996kz}. It performs robust linear regression on data while accounting for measurement uncertainties in both the $x-$ and $y-$ directions, intrinsic scatter, and correlations between errors, although it can be sensitive to outliers and certain assumptions. In the following sections, we assess the performance of the method in determining the best-fit values of $\gamma$ and $\beta$. We note, however, that for this quasar sample, the substantial dispersion in the data prevents the method from yielding reliable results. Therefore, we adopt an alternative procedure based on a Bayesian approach, implemented with \textsc{Linmix} \cite{Kelly:2007jy}. \textsc{Linmix} performs Bayesian linear regression accounting for measurement uncertainties in both variables and an intrinsic-scatter term, and it samples the posterior of the calibration parameters via MCMC; we monitor convergence using standard diagnostics. The parameter $\rho$ reported by \textsc{Linmix} corresponds to the correlation coefficient between the underlying (error-deconvolved) regression variables, here $x\equiv \log L_{\rm UV}$ and $y\equiv \log L_{\rm X}$.

%%%%%%%%%%%%%%%%%%%%%%%%%%%%%%%%%%%%%%%%%%%%%%%%%%%%%%%%%%%%%%%%%%%%%%%%%%%%%%%%%%%
%%%%%%%%%%%%%%%%%%%%%%%%%%%%%%%%%%%%%%%%%%%
\section{Dark Energy models} 
\label{Sec:III}
%%%%%%%%%%%%%%%%%%%%%%%%%%%%%%%%%%%%%%%%%%%

The dominance of dark energy drives a phase of accelerated cosmic expansion, which can be effectively characterized by its equation of state parameter $w_{\rm DE}=p_{\rm DE}/\rho_{\rm DE}$, where the subscript ${\rm DE}$ refers to dark energy. Within the framework of a homogeneous and isotropic Universe described by the FLRW metric, accelerated expansion occurs when the pressure is sufficiently negative, specifically when $w_{\rm DE} < -1/3$. This parameter not only governs the gravitational behavior of dark energy but also dictates its dynamical evolution through the conservation of the energy-momentum tensor. Under the assumptions of spatial flatness, pressureless matter, and negligible radiation, the Friedmann equation simplifies to:  
\begin{equation}
    \frac{H^2(z)}{H^2_0}= \Omega_m (1+z)^3 + \Omega_{\rm DE} \exp{\left(3 \int \frac{dz'}{1+z'}[1+w_{\rm DE}(z')]\right)},
    \label{DE_models}
\end{equation}

\noindent where the density fraction parameters are defined as $\Omega_m\equiv \rho_m(t_0)/\rho_c^0$ and $\Omega_{\rm DE}\equiv \rho_{\rm DE}(t_0)/\rho_c^0$ with critical density $\rho_c^0\equiv 3H_0^2/(8 \pi G)$.

From Eq. (\ref{DE_models}), we recover the expansion history for the dark energy models studied here: 
\begin{enumerate}
\item $\Lambda$CDM model
    
In this case Eq. \eqref{DE_models} reads,
    \begin{equation}
    H^2(z)= H^2_0\left[\Omega_m (1+z)^3 + \Omega_{\rm DE}\right].
    \label{LCDM_model}
\end{equation}
Here, $\Omega_{\Lambda}\equiv \Omega_{\rm DE}$ is the density parameter associated with a cosmological constant, characterized by an equation of state $w_{\Lambda}=-1$. Imposing consistency of Eq. \eqref{LCDM_model} at $z=0$, that is, $H(z=0)=H_0$, leads to the condition $\Omega_m+\Omega_{\Lambda}=1$.

\item $w$CDM model
    
For the case in which $w_{\rm DE}$ is a constant such that $w_{\rm DE} \equiv w \ne -1$, one gets
    \begin{equation}
    H^2(z)= H^2_0\left[\Omega_m (1+z)^3 + \Omega_{\rm DE}(1+z)^{3(1+w_{\rm DE})}\right],
    \label{wcdm_model}
\end{equation}
where $\Omega_{\rm DE}$ is the density fraction due to the dark energy fluid. Given that spatial flatness requires $\Omega_m+\Omega_{\rm DE}=1$, the free parameters of this model reduce to $\Omega_m$ and $w_0$.

\item CPL model
    
In the widely used CPL parametrization, the dark-energy equation of state is written as $w_{\rm DE} = w_0 + w_a \frac{z}{1+z}$, with $w_0$ and $w_a$ taken as constants. Using Eq. \eqref{DE_models}, the corresponding Hubble parameter is
   \begin{equation}
    H^2(z)= H^2_0\left[\Omega_m (1+z)^3 + \Omega_{\rm DE}(1+z)^{3(1+w_0+w_a)} \exp{\left( -\frac{3w_a z}{1+z}\right)}\right],
    \label{cpl_model}
\end{equation}
where $\Omega_{\rm DE}$ denotes the present-day dark-energy density fraction, and we impose spatial flatness so that $\Omega_m+\Omega_{\rm DE}=1$. The model is therefore specified by the three free parameters $\Omega_m$, $w_0$ and $w_a$.
\end{enumerate}

In the following section, we proceed to evaluate the reliability of quasars as cosmological probes for constraining the free parameters of these cosmological models through Bayesian parameter estimation. To this end, we use a set of up-to-date cosmological observations, including Type Ia Supernovae, CMB data in the form of shift parameters, the latest DESI DR2 BAO measurements, and the calibrated quasar sample presented here.

%%%%%%%%%%%%%%%%%%%%%%%%%%%%%%%%%%%%%%%%%%%%%%%%%%%%%%%%%%%%%%%%%%%%%%%%%%%%%%%%%%%%%%%%
\section{Computational tools and observational samples}
\label{Sec:IV}

We employ the public Boltzmann code \textsc{CLASS} \citep{Lesgourgues2011} to compute the background evolution for all dark-energy models considered in this work. 
In the stepwise analysis (calibration followed by cosmological inference with fixed calibration parameters), parameter estimation is performed with \textsc{MontePython} \citep{Audren2013}, interfaced with \textsc{CLASS}, using a Metropolis–Hastings MCMC sampler \citep{Metropolis:1953am,Hastings:1970aa}. Convergence is assessed through the Gelman–Rubin criterion \citep{gelman1992}, requiring $R-1 < 10^{-3}$ for all runs.

For the joint calibration–cosmology inference, we sample the posterior distribution with the \textsc{emcee} MCMC sampler \citep{Foreman-Mackey_2013}. Following the prior choices described in Section \ref{sec:priors}, we fit the cosmological parameters simultaneously with the quasar-calibration parameters $\{\beta,\gamma,\delta \}$, and we additionally vary the supernova nuisance parameter $M_{\rm SN}$ when supernova data are included. The sampled cosmological parameters are $\{\Omega_m, H_0\}$ for $\Lambda$CDM, $\{\Omega_m, w, H_0\}$ for $w$CDM, and $\{\Omega_m, w_0, w_a, H_0\}$ for CPL.

The same observational probes are used whenever the same dataset combinations are considered. We note, however, that the implementation of the SH0ES information differs between the stepwise and joint analyses: in the MontePython stepwise runs we use the public \texttt{Pantheon\_Plus\_SH0ES} likelihood, which includes Cepheid-calibrator information, whereas in the joint \texttt{emcee} analysis SH0ES is included as an external Gaussian likelihood on $H_0$.

In addition to the calibrated quasar samples discussed above, we incorporate a suite of complementary datasets that probe the expansion history entering Eq.~(\ref{DE_models}). These measurements trace the distance–redshift relation, and are described in the following subsections.

\subsection{Priors and sampling strategy}
\label{sec:priors}

Because two different inference strategies are adopted in this work (stepwise and joint calibration–cosmology), we explicitly report the priors used in each case.

\paragraph*{Stepwise analysis (MontePython).}

\begin{table}
\caption{Priors adopted in the stepwise \textsc{MontePython} analysis. These priors apply only to the configuration where the quasar calibration parameters are fixed. When the SN+SH0ES likelihood is included, the $H_0$ information is provided by that likelihood term; we do not apply an additional independent $H_0$ prior on top of it.}
\label{tab:priors}
\centering
\begin{tabular}{ll}
\hline
Parameter & Prior \\
\hline
\multicolumn{2}{c}{\textbf{$\Lambda$CDM}} \\
$\Omega_{\rm m}$ & flat $[0.10,\,0.99]$ \\
$H_0$ & flat $[20,\,100]$ \\
\hline
\multicolumn{2}{c}{\textbf{$w$CDM}} \\
$\Omega_{\rm m}$ & flat $[0.10,\,0.99]$ \\
$w$ & flat (unbounded) \\
$H_0$ & flat $[20,\,100]$ \\
\hline
\multicolumn{2}{c}{\textbf{CPL}} \\
$\Omega_{\rm m}$ & flat $[0.10,\,0.99]$ \\
$w_0$ & flat (unbounded) \\
$w_a$ & flat $[-9,\,2]$ \\
$H_0$ & flat $[20,\,100]$ \\
\hline
\end{tabular}
\end{table}

In the stepwise configuration, cosmological parameters are sampled with \textsc{MontePython} interfaced with \textsc{CLASS}. We adopt broad uniform (flat) priors to define a physically reasonable parameter domain when the quasar calibration parameters are kept fixed. These priors are adopted for numerical stability and efficient sampling in the stepwise setup and are not intended to add external cosmological information beyond the datasets explicitly included.

When the SN+SH0ES dataset is included, the $H_0$ information is provided by the SH0ES likelihood term itself; we do not add any extra independent $H_0$ prior on top of that.

For all stepwise runs we adopt identical priors and initialization settings to ensure internal consistency across datasets and models (see Table~\ref{tab:priors}). In the \textsc{CLASS} computations, cosmological parameters not varied in a given model (e.g. $n_s$, $\ln(10^{10}A_s)$, $\tau$, $\Omega_b$, and neutrino-sector parameters) are fixed to the Planck 2018 best-fit values 
\citep{Aghanim:2018eyx}, since the present analysis focuses exclusively on late-time background expansion.

\paragraph*{Joint calibration–cosmology analysis (\textsc{emcee}).}

\begin{table}
\caption{Priors adopted in the joint calibration–cosmology analysis (\textsc{emcee}). 
For each cosmological model we sample only the corresponding subset of parameters.}
\label{tab:priors_emcee}
\centering
\begin{tabular}{lll}
\hline
Parameter & Prior type & Range \\
\hline
$\Omega_m$ & flat & [0.01, 0.99] \\
$H_0$ & flat & [20, 100] \\
$w$ & flat & [-2, 0] \\
$w_0$ & flat & [-2, 0] \\
$w_a$ & flat & [-9, 2] \\
$\beta$ & flat & [4, 10] \\
$\gamma$ & flat & [0.4, 0.8] \\
$\delta$ & flat & [0, 1] \\
$M_{\rm SN}$ & flat & $[-30,-10]$ \\
\hline
\end{tabular}
\end{table}

In the joint inference framework, cosmological and quasar-calibration parameters are sampled simultaneously using \textsc{emcee}. Table~\ref{tab:priors_emcee} lists the prior ranges used in the joint runs; for each cosmological model we sample only its corresponding subset of parameters. We adopt uniform (flat) priors on all free parameters. For each cosmological scenario, the sampled subset is $(\Omega_m, H_0)$ for $\Lambda$CDM, $(\Omega_m, w, H_0)$ for $w$CDM, and $(\Omega_m, w_0, w_a, H_0)$ for CPL, together with the calibration parameters $(\beta,\gamma,\delta)$ (and $M_{\rm SN}$ when supernovae are included).

The prior ranges are chosen to be broad and physically motivated, while remaining consistent with the parameter ranges adopted in previous quasar calibration and late-time cosmological analyses. For the QSO calibration parameters, the posterior distributions are well contained within the prior boundaries, indicating that these constraints are driven by the data. For parameters associated with the absolute distance scale, however, the interpretation depends on the analysis case. In particular, when the SH0ES information is removed, $H_0$ becomes weakly constrained and can exhibit broad, non-Gaussian posteriors reflecting its degeneracy with $M_{\rm SN}$. In the CPL model, the $w_a$ posterior can also be broad or prior-affected. We therefore report credible intervals rather than point estimates whenever the posterior is strongly non-Gaussian or prior-limited.

The same hard prior ranges listed in Table~\ref{tab:priors_emcee} are supplied to GetDist when producing the marginalized posterior plots. This ensures that posteriors affected by prior boundaries, such as the broad $H_0$ posteriors in Case C and the poorly constrained $w_a$ distributions in CPL, are displayed with the correct prior edges.

The only informative external constraint considered in the joint framework is the SH0ES $H_0$ information in Case A, implemented as a Gaussian likelihood term. In Case C no external $H_0$ information is included.

\subsection{Type Ia Supernovae (SNe Ia)}

In this study, we use the Pantheon+ Type Ia supernova sample
\citep{Scolnic_2022,Brout:2022vxf}, which comprises 1701
light curves from 1550 spectroscopically confirmed SNe Ia spanning
the redshift range $0.001<z<2.26$. The Pantheon+ sample provides
apparent magnitudes rather than distance moduli, and the supernova
likelihood is computed from apparent-magnitude residuals using the
statistical-plus-systematic covariance matrix provided with the
Pantheon+/SH0ES release.

In the stepwise MontePython analysis, when SH0ES information is
included, we use the public \texttt{Pantheon\_Plus\_SH0ES} likelihood,
which incorporates the Cepheid-calibrator information through the
calibrator entries in the Pantheon+ release. In this implementation,
Cepheid-host calibrators are compared against their Cepheid distances,
whereas the remaining SNe are compared against the cosmological
distance predicted by CLASS.

In the joint calibration--cosmology analysis, instead, we use the
Pantheon+ supernova likelihood and include the SH0ES information
as an external Gaussian likelihood on $H_0$,

\begin{equation}
\ln \mathcal{L}_{\rm SH0ES}
=
-\frac{1}{2}
\left(\frac{H_0-73.04}{1.04}\right)^2
-\ln(1.04)-\frac{1}{2}\ln(2\pi),
\end{equation}
corresponding to $H_0=73.04\pm1.04~{\rm km\,s^{-1}\,Mpc^{-1}}$. Thus, in the joint analysis, cases labeled SN+SH0ES include the
Pantheon+ supernova likelihood, supplemented by this external SH0ES
$H_0$ likelihood term. No additional independent $H_0$ prior is
imposed beyond the broad flat sampling prior.

%%%

\subsection{Baryon Acoustic Oscillations (BAO)}

We use the latest Dark Energy Spectroscopic Instrument (DESI) Data Release 2 (DR2) BAO sample \cite{DESI2025arXiv250314738D} (hereafter DESI). 

The recent DESI dataset is built from observations of multiple tracers, including bright galaxy samples (BGS), luminous red galaxies (LRGs), emission line galaxies (ELGs), quasars, and the Ly$\alpha$ forest, spanning the redshift range $0.1 < z<4.2$. Accordingly, we make use of the measurements of the comoving distance, $D_M(z)/r_d$, and the Hubble distance, $D_H(z)/r_d$, where $r_d$ denotes the sound horizon at the drag epoch and
\begin{equation}
    D_M(z)\equiv \int_0^z \frac{cdz'}{H(z')}, \quad D_H(z)\equiv\frac{c}{H(z)}.
\end{equation}
In addition, we include the angle-averaged distance measure $D_V(z)/r_d$, defined as 
\begin{equation}
    D_V \equiv [z D_M(z)^2D_H(z)]^{1/3}.
\end{equation}

The BAO distance measurements from DESI DR2 are reported in Table IV in \cite{DESI2025arXiv250314738D}. Since the public DESI BAO likelihood is compatible with the Cobaya sampler, it was adapted in \cite{Herold:2025hkb} for use with MontePython, which is the version we employed in this work.

\subsection{Cosmic Microwave Background (CMB)}

Rather than using the full CMB anisotropy likelihood, we adopt the condensed form of CMB information provided by the shift parameters reported by \cite{Chen:2018dbv}, derived from the final Planck release \citep{Aghanim:2018eyx}, hereafter \textit{Planck compressed}. This approach significantly reduces the computational cost while retaining the late-time geometric information of the CMB that is most relevant for background expansion constraints. Several studies have shown that the shift parameters $(R,l_A,\Omega_bh^2,n_s)$ provide an efficient and accurate summary of CMB information for constraining dark-energy models \citep{Kosowsky:2002zt,Wang:2007mza,Mukherjee:2008kd,Ade:2015rim}, particularly for extensions of the standard $\Lambda$CDM framework with a smooth dark-energy component, as considered here. Since our analysis focuses exclusively on late-time background parameters (e.g. $\Omega_m$, $H_0$, $w$, $w_0$, $w_a$) rather than on primordial perturbation, or foreground nuisance parameters, employing the full Planck likelihood would substantially enlarge the parameter space and increase runtime without providing additional constraining power for the quantities of interest.

The first two quantities in the vector $(R,l_A,\Omega_bh^2,n_s)$ are defined as 
\begin{equation}
R \equiv \sqrt{\Omega_m H_0^2} \frac{r(z_{*})}{c},
\end{equation}
\begin{equation}
l_A\equiv \pi \frac{r(z_{*})}{r_s(z_{*})},
\end{equation}
where $r(z)$ is the comoving distance, here evaluated at photon-decoupling epoch $z_{*}$. 
The corresponding $\chi^2$ for the CMB is thus
\begin{equation}
\chi^2_{CMB}= \mathrm{\Delta} \mathcal{F}^{CMB} \cdot \mathbf{C}_{CMB}^{-1} \cdot \mathrm{\Delta} \mathcal{F}^{CMB}    ,
\label{Eq:chiCMB}
\end{equation}
where $\mathcal{F}^{CMB}=(R,l_A,\Omega_bh^2,n_s)$ is the vector of the shift parameters and $\mathbf{C}^{-1}_{CMB}$ is the respective inverse covariance matrix. The mean values for these shift parameters as well as their standard deviations and normalized covariance matrix are taken from Table 1 of \cite{Chen:2018dbv}. 

Before proceeding, we emphasize that, because of the $H_0$ tension, any analysis in this work that applies both SH0ES and CMB distance priors simultaneously is included only as a diagnostic benchmark and is not interpreted as a definitive combined constraint.

%%%%%%%%%%%%%%%%%%%%%%%%%%%%%%%%%%%%%%%%%%%%%%%%%%%%%%%%%%%%%%%%%%%%%%%%%%%%%%%%%%%%%%%%%%

\section{Calibration results and cosmological constraints}
\label{Sec:V}

In this section we present the results obtained from the quasar calibration and their implications for cosmological parameter estimation. 
We first describe a stepwise strategy, in which the calibration is performed independently and subsequently used for cosmological inference. 
We then introduce a joint calibration–cosmology analysis, where both sets of parameters are inferred simultaneously within a single likelihood.

\subsection{Performance of \textsc{BCES} and \textsc{Linmix} Algorithms in Quasar Calibration}
Before performing any cosmological inference, we first assess the robustness of the quasar calibration itself. We compare two commonly used regression approaches, \textsc{BCES} and the Bayesian method \textsc{Linmix}, in order to determine which provides more reliable estimates of the $L_X$–$L_{UV}$ relation.

For the full dataset of 2038 quasars spanning $0.009<z<7.541$, the \textsc{BCES} fit to Eq.~\eqref{Eq:nonlinear} yields $\gamma=0.690 \pm 0.008$ and $\beta=5.656 \pm 0.251$. These values differ noticeably from those reported in previous analyses \citep{Lusso:2020pdb,Lusso:2025bhy,Lusso2017AA}, where $\gamma \sim 0.6$ and $\beta \sim 8$. To further investigate this discrepancy, we also apply \textsc{BCES} to the binned version of the sample, with the results summarized in Table~\ref{tab:BCES values}.

\begin{table}
\caption{Best-fit of the quasar parameters and their $1\sigma$ uncertainties for the binned version of the dataset using the \textsc{BCES} algorithm.}
\label{tab:BCES values}
\begin{center}
\begin{tabular}{c  c  c }
\hline
 Bin & $\gamma$ &  $\beta$ \\ \hline
0 & 0.572 $\pm$ 0.099 &  9.054 $\pm$ 2.905 \\
1 &  0.574 $\pm$ 0.039 &   9.122 $\pm$ 1.178 \\
2 &  0.621 $\pm$ 0.036 & 7.742 $\pm$ 1.087 \\
3 & 0.545 $\pm$ 0.028 & 10.104 $\pm$ 0.861 \\
4 &  0.613$\pm$ 0.033& 8.054 $\pm$ 1.0133 \\
5 &  0.519 $\pm$0.028 & 11.053 $\pm$ 0.869 \\
6 & 0.600 $\pm$ 0.032 & 8.589 $\pm$ 1.019  \\
7 &  0.583 $\pm$ 0.047 & 9.162 $\pm$ 1.508  \\
8 &  0.547$\pm$ 0.094 & 10.411 $\pm$ 3.039 \\
9 &  0.574 $\pm$ 0.568 & 9.551 $\pm$ 18.283 \\
10 &  0.790 $\pm$ 1.927 & 2.640 $\pm$ 62.056 \\ \hline
\end{tabular}
\end{center}
\end{table}

As shown in Table~\ref{tab:BCES values}, \textsc{BCES} exhibits instabilities in bins with sparse data, where the uncertainties increase significantly. This behavior is consistent with known limitations of the method when applied to small samples with large measurement errors \citep{Kelly:2007jy}. In addition, \cite{Andreon2013} noted that \textsc{BCES} can be sensitive to outliers, non-Gaussian distributions, upper limits, and selection effects. Another limitation concerns the treatment of intrinsic scatter, which in the standard \textsc{BCES} formulation is only properly defined when the independent variable is measured without error, a condition not satisfied in our case.

Given these considerations, we adopt the Bayesian regression algorithm \textsc{Linmix} \citep{Kelly:2007jy}. This method uses MCMC sampling to infer the full posterior distributions of the regression parameters and naturally accounts for measurement uncertainties in both variables, intrinsic scatter, and selection effects. Although computationally it is more demanding, it generally provides more stable and robust estimates for heterogeneous astronomical datasets.

Table~\ref{tab:LinmixValues} presents the \textsc{Linmix} results for the binned sample, allowing a direct comparison with \textsc{BCES}. While both methods give broadly consistent estimates in the lowest-redshift bins, noticeable differences appear in bins with limited statistics, where \textsc{Linmix} remains more stable. We emphasize that the binned fits are used only for diagnostic purposes and are not employed in the cosmological analysis, since binning can smooth intrinsic scatter and bias the inferred parameters.

\begin{table}
\caption{Best-fit of the quasar parameters and their $1\sigma$ uncertainties for the binned version of the dataset using the \textsc{Linmix} algorithm.}
\label{tab:LinmixValues}
\centering
\begin{tabular}{cccccc}
\hline
$\gamma$ & $\beta$ & $\sigma_{\gamma \beta}$ & $\delta$ & $\rho$ \\
\hline
$0.541 \pm 0.076$ & $9.961 \pm 2.220$ & -0.169 & $0.221 \pm 0.066$ & 0.924 \\
$0.571 \pm 0.039$ & $9.208 \pm 1.174$ & -0.046 & $0.237 \pm 0.011$ & 0.695 \\
$0.620 \pm 0.036$ & $7.793 \pm 1.099$ & -0.040 & $0.251 \pm 0.010$ & 0.707 \\
$0.539 \pm 0.030$ & $10.272 \pm 0.920$ & -0.027 & $0.219 \pm 0.009$ & 0.708 \\
$0.611 \pm 0.030$ & $8.123 \pm 0.926$ & -0.028 & $0.229 \pm 0.009$ & 0.746 \\
$0.515 \pm 0.024$ & $11.190 \pm 0.766$ & -0.019 & $0.201 \pm 0.009$ & 0.777 \\
$0.590 \pm 0.029$ & $8.898 \pm 0.926$ & -0.027 & $0.159 \pm 0.011$ & 0.852 \\
$0.581 \pm 0.046$ & $9.248 \pm 1.465$ & -0.067 & $0.190 \pm 0.021$ & 0.856 \\
$0.494 \pm 0.115$ & $12.142 \pm 3.704$ & -0.424 & $0.122 \pm 0.060$ & 0.919 \\
$0.590 \pm 0.391$ & $9.047 \pm 12.595$ & -4.928 & $0.291 \pm 0.143$ & 0.676 \\
$0.496 \pm 0.173$ & $12.224 \pm 5.574$ & -0.962 & $0.240 \pm 0.096$ & 0.784 \\
\hline
\end{tabular}
\end{table}

\begin{table*}
\caption{\label{tab:linmix2}
Best-fit quasar parameters and their associated $1\sigma$ uncertainties, obtained using \textsc{Linmix} algorithm, for the full sample (Quasars all $z$-range), the low-redshift subsample ($z\leq1.43$, Quasars low-$z$) and the high-redshift subsample ($z>1.43$, Quasars high-$z$). }
%\footnotetext[3]{low-z: 972 quasars}
%\footnotetext[4]{high-z: 1066 quasars}
\begin{tabular}{cccc}
\hline
 &Quasars all z-range &Quasars low-z ($z<1.43$)&Quasars high-z ($z>1.43$)\\
\hline
$\gamma$ & $0.688 \pm 0.009$  & $0.618 \pm 0.017$ & $0.632 \pm 0.013$  \\
$\beta$ & $5.747 \pm 0.264$ & $7.824 \pm 0.522$ & $7.515 \pm 0.412$ \\
$\delta$   & $0.231\pm 0.004$& $0.237\pm 0.006$ & $0.211\pm 0.006$ \\
$\rho$   & $0.89 \pm 0.01$ & $0.77 \pm 0.01$ & $0.86 \pm 0.01$ \\
\hline
\end{tabular}
\end{table*}

%%%%%%%%%figure%%%%%%%%%%%%%%%
\begin{figure}
   \centering

   \includegraphics[width=3.5 in]{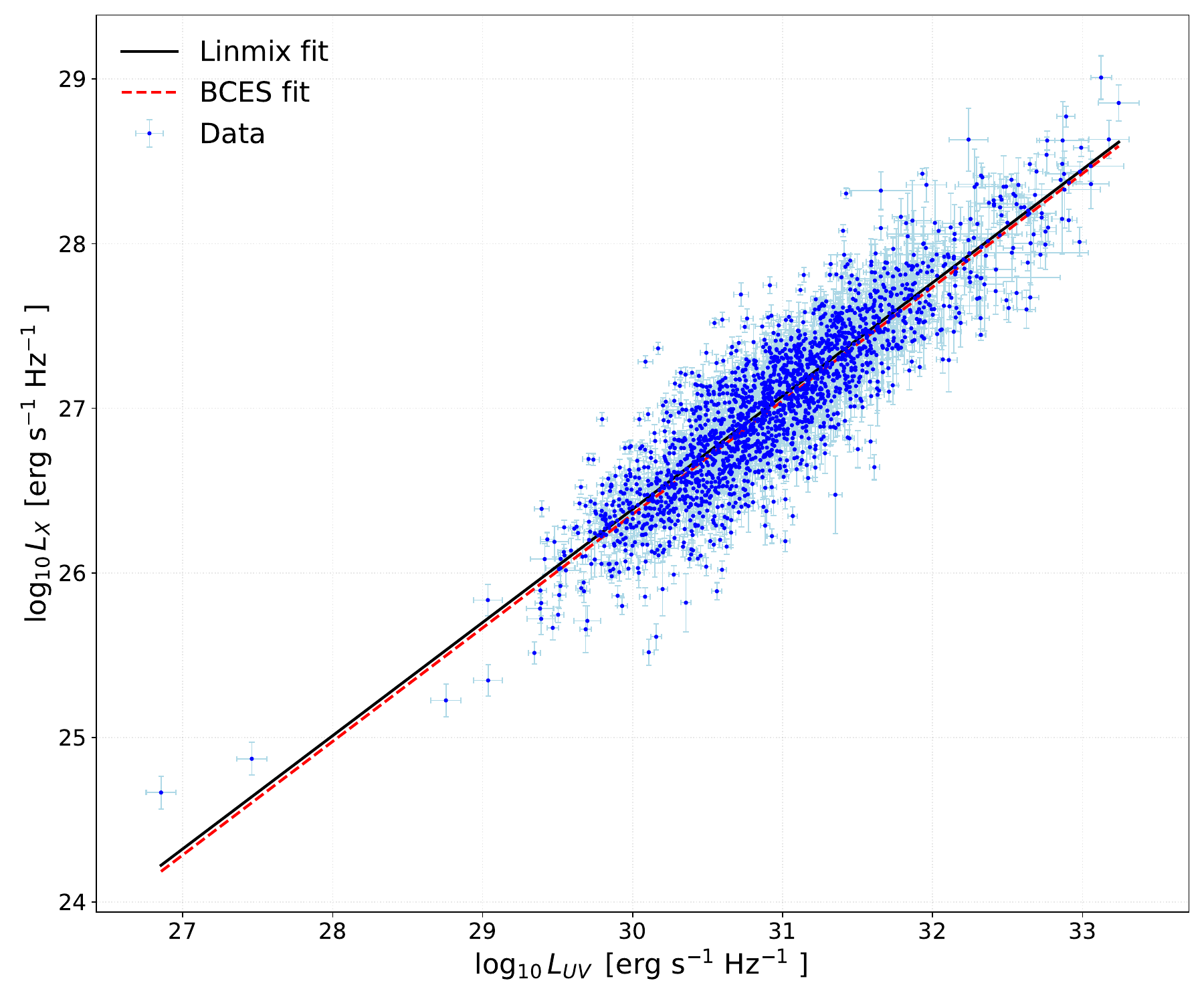}
    \caption{Rest-frame monochromatic luminosities $\log(L_X)$ against $\log(L_{UV})$ for the final sample of 2038 quasars (blue circles) as described in Sect. 2.1. The results from the BCES fit (dashed red line) and from the Linmix fit (black solid line) are also reported.  }
   \label{ref:LxLuv_plot}
\end{figure}
%%%%%%%%%%%%%%%%%%%%%%%%%%%%%%%

All subsequent calibrations therefore rely on \textsc{Linmix} applied to the unbinned data. We perform two configurations: (i) the full sample and (ii) separate fits to low and high redshift subsamples. The resulting best-fit parameters are listed in Table~\ref{tab:linmix2}, and Fig.~\ref{ref:LxLuv_plot} shows the corresponding $L_X$–$L_{UV}$ relation together with the \textsc{BCES} fit for reference.

For the low-redshift quasars ($z\leq 1.43$), \textsc{Linmix} yields $\gamma=0.618 \pm 0.017$ and $\beta=7.824 \pm 0.522$, consistent within $1\sigma$ with the values reported by \cite{Lusso:2025bhy}. This agreement supports the reliability of the calibration in this redshift range and motivates the use of this sample in the subsequent cosmological analysis.

\subsection{Hubble diagram}

Using the calibration parameters obtained in the previous subsection, we compute the quasar distance modulus as  $\mu_{\mathrm{quasar}}=5 \log(d_L^{\mathrm{cal}}/\mathrm{Mpc}) +25 $, where the calibrated luminosity distance is derived from 
\begin{equation} 
\log(F^{\mathrm{cal}}_X) = \gamma \log(F_{UV}) + (2\gamma - 2)\log(d^{\mathrm{cal}}_L) + (\gamma - 1)\log(4\pi) + \beta.
\end{equation} 

The values of $\gamma$ and $\beta$ depend on whether the fit is performed using the full dataset or restricted to the low or high redshift subsamples.

The uncertainty on $\mu_{\mathrm{quasar}}$ is estimated through standard error propagation, accounting for the covariance between the calibration parameters as well as the measurement uncertainties in $F_X$ and $F_{UV}$ as follows,
\begin{align}
\begin{split}
    \sigma_{\mu_{\mathrm{quasar}}}^2=&\left(\frac{\partial \mu_{\mathrm{quasar}}}{\partial \gamma}\right)^2\sigma_{\gamma}^2 + \left(\frac{\partial \mu_{\mathrm{quasar}}}{\partial \beta}\right)^2\sigma_{\beta}^2 \\ 
    & + 2\left(\frac{\partial \mu_{\mathrm{quasar}}}{\partial \gamma}\right)\left(\frac{\partial \mu_{\mathrm{quasar}}}{\partial \beta}\right)\sigma_{\gamma \beta}\\
    & +\left(\frac{\partial \mu_{\mathrm{quasar}}}{\partial S_{\mathrm{Log(F_X)}}}\right)^2\sigma_{\mathrm{Log(F_X)}}^2+\left(\frac{\partial \mu_{\mathrm{quasar}}}{\partial S_{\mathrm{Log(F_{UV})}}}\right)^2\sigma_{\mathrm{Log(F_UV)}}^2.
\end{split}
\end{align}

%%%%%%%%%figure%%%%%%%%%%%%%%%
\begin{figure}
   \centering

   \includegraphics[width=3.5 in]{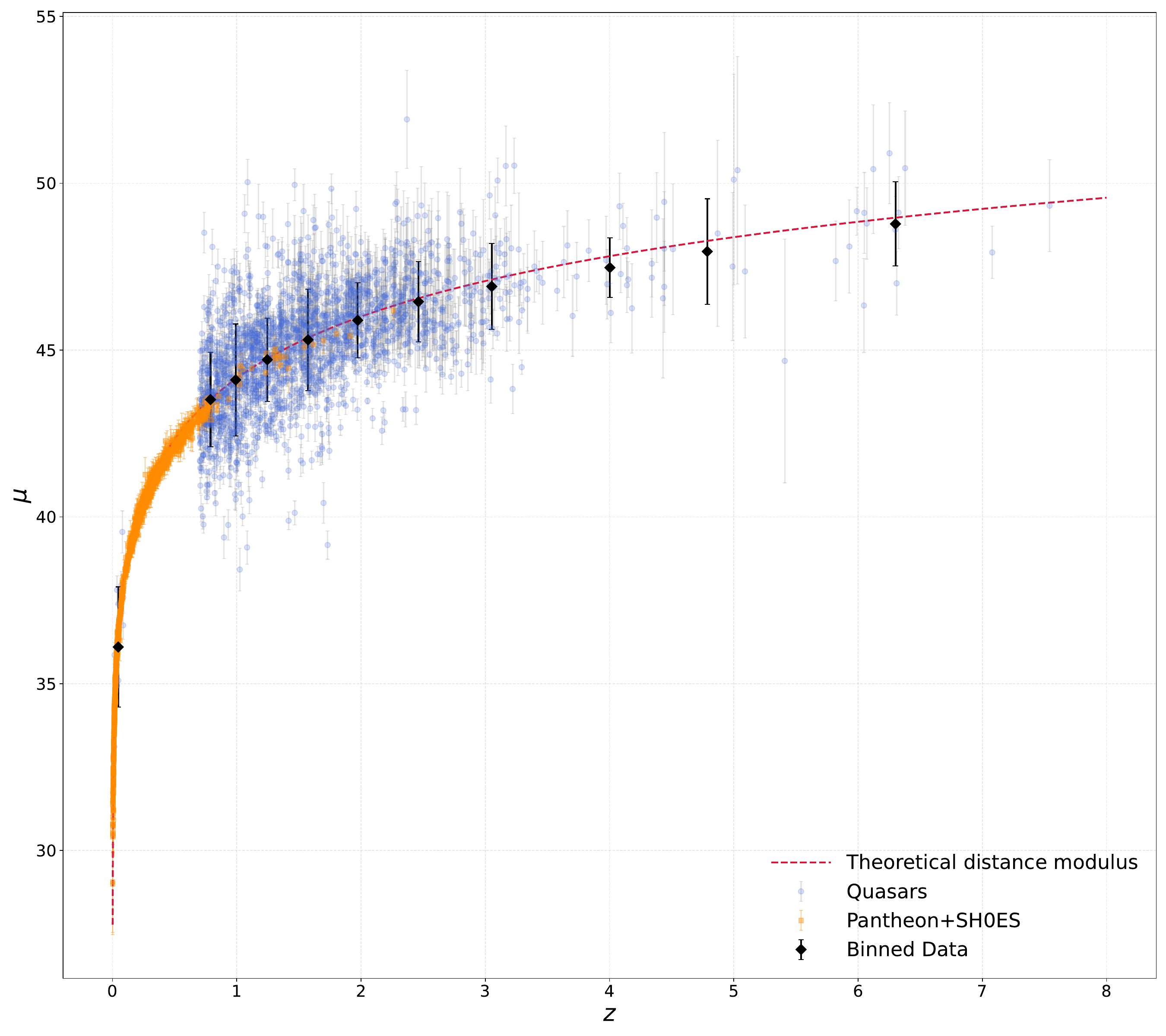}
    \caption{Hubble diagram of Pantheon+ supernovae (orange
points) and quasars in the redshift range $0.009<z<7.541$ (blue points). The black points are the binned quasar DM. The dotted line in red is the flat $\Lambda$CDM model with $\Omega_m=0.315$ and $H_0= 67.66$ km/s/Mpc.}
   \label{ref:HubbleDiagram}
\end{figure}
%%%%%%%%%%%%%%%%%%%%%%%%%%%%%%%

Figure~\ref{ref:HubbleDiagram} shows the resulting Hubble diagram combining quasars and Pantheon+ supernovae. The red dotted curve corresponds to the prediction of a reference flat $\Lambda$CDM model with $\Omega_m = 0.315$ and $H_0 = 67.66$ km s$^{-1}$ Mpc$^{-1}$ \citep{Aghanim:2018eyx}, following 
\begin{equation}
    d^{\Lambda\mathrm{CDM}}_{L}(z)=c(1+z)\int_{0}^{z} \frac{dz'}{H_0 \sqrt{\Omega_m(1+z')^3 + (1-\Omega_m)}}.
    \label{Ec:dL_LCDM}
\end{equation}

Overall, the quasar distance moduli follow the expected trend of the standard cosmological model across the full redshift range, extending the Hubble diagram well beyond the supernova regime. This agreement provides a qualitative validation of the calibration and motivates the use of quasars as high-redshift distance indicators.

A quantitative assessment of the cosmological constraints derived from these calibrated samples is presented in the following subsection.

%%%%%%%%%%%%%%%%%%%%%%%%%%%%%%%%%%%%%%%%%%%%%%%%%%%%%%%%%%%%%%%%%%%%%%%%%%%%%%%

\subsection{Stepwise cosmological constraints}

We first adopt the traditional stepwise strategy, in which the quasar calibration parameters are fixed to their best-fit values and subsequently used to constrain cosmological models. This approach serves as a useful baseline and allows direct comparison with previous studies. However, because the calibration and cosmological parameters are not varied within a single likelihood, the stepwise constraints should be interpreted as diagnostic benchmarks rather than as our main cosmological results. Our main inference framework is the joint calibration--cosmology analysis presented later, where these correlations are propagated by construction.

As a consistency check of the standard datasets used in this work,
we first derive constraints using the public
\texttt{Pantheon\_Plus\_SH0ES} likelihood (SN+SH0ES)  \citep{Brout:2022vxf,Scolnic_2022}. We then consider SN+SH0ES in combination with the DESI DR2 BAO measurements \citep{DESI2025arXiv250314738D} and with the Planck compressed CMB distance priors \citep{Chen:2018dbv}. After establishing these baseline constraints, we include the calibrated quasar sample and repeat the inference for each cosmological scenario, using both the full quasar sample (up to $z\simeq7.5$) and the low-redshift subsample ($z\leq1.43$).

It is worth emphasizing a technical point. In the stepwise approach, the quasar calibration is anchored to external $H(z)$ information, which fixes the absolute distance scale of the calibrated quasar Hubble diagram. Once the normalization parameter of the quasar relation is fixed, this absolute scale is also fixed. Since this scale is degenerate with $H_0$, the stepwise approach cannot self-consistently readjust the quasar calibration and the cosmological parameters within a single likelihood. Therefore, we do not report 
$H_0$ constraints for QSO-inclusive stepwise combinations. A self-consistent treatment of \(H_0\) requires leaving the absolute calibration free and sampling it together with the cosmological parameters, as done in the joint calibration--cosmology analysis presented in the next subsection.

For the stepwise analysis, we present the constraints from the full QSO sample combined with SNe as the main diagnostic case, since it provides the highest statistical leverage within this configuration. The corresponding results for the low-\(z\) subsample are reported in the tables as a robustness check, but we focus our figures on the full-sample constraints for clarity.

We stress that all QSO-inclusive stepwise constraints reported in Tables~\ref{tab:lcdm}, \ref{tab:wcdm}, and \ref{tab:cpl} should be read as diagnostic results. They are not used as our main cosmological constraints. In these cases, the quasar calibration is anchored to the CC-based Bézier reconstruction of $H(z)$, whereas the subsequent cosmological fits assume the expansion history of the specific model under consideration, namely $\Lambda$CDM, $w$CDM, or CPL. Therefore, the stepwise procedure does not guarantee that the $H(z)$ used in the calibration stage coincides with the $H(z)$ preferred in the cosmological-inference stage. Moreover, combinations involving both SH0ES and CMB distance priors are kept only as diagnostic benchmarks, given the known tension between these datasets. Our conclusions do not rely on interpreting these combinations as statistically definitive joint constraints.

\subsubsection{Results for $\Lambda$CDM}

\begin{table*}
\caption{\label{tab:lcdm}
\textbf{Diagnostic stepwise constraints} for the flat $\Lambda$CDM parameters obtained in the stepwise analysis using SN+SH0ES, DESI DR2 BAO (DESI), Planck compressed CMB distance priors, and calibrated QSOs (Full QSO Sample and QSO low-$z$, $z<1.43$).  }
\begin{tabular}{lccc}
\hline
 &$\Omega_m$ & $H_0$ \\
\hline 
SN+SH0ES & $0.333^{+0.019}_{-0.020} $   & $73.72 \pm 1.000$  \\ 
SN+SH0ES + DESI  & $0.313^{+0.008}_{-0.009}$ & $71.14 ^{+0.700}_{-0.740}$ \\ 
SN+SH0ES + Planck Compressed & $0.288 \pm 0.002$ & $69.42 \pm 0.200$ \\ 
SN+SH0ES + DESI  + QSO low-$z$&  $0.630 \pm 0.007$ &  - \\ 
SN+SH0ES + DESI  + Full QSO Sample & $0.538\pm 0.006$&  - \\ 

SN+SH0ES + Planck Compressed + QSO low-$z$&  $0.280 \pm 0.001$ &  - \\ 
SN+SH0ES + Planck Compressed + Full QSO Sample & $0.280\pm 0.001$&  - \\ 

\hline
\end{tabular}
\end{table*}

\begin{table}
\begin{center}
\caption{\label{tab:lcdm_PPSQuasars}
Diagnostic stepwise constraints for the flat $\Lambda$CDM model derived using SN+SH0ES in combination with the calibrated QSOs (Full QSO Sample and QSO low-$z$, $z<1.43$).}
\begin{tabular}{lccc}
\hline
 &$\Omega_m$ \\
\hline 
SN+SH0ES+ QSO low-$z$& $0.718 \pm 0.009$  \\
SN+SH0ES  + Full QSO Sample & $0.592\pm 0.007$  \\

\hline
\end{tabular}
\end{center}
\end{table}

Table~\ref{tab:lcdm} summarizes the diagnostic stepwise constraints for the flat $\Lambda$CDM model. For the non-QSO reference cases, the recovered parameter values are consistent with expectations from the corresponding likelihood combinations and confirm that the likelihoods and sampling pipeline behave as expected. See, for example, Table 3 in \cite{Brout:2022vxf}.

When the calibrated QSO samples are incorporated within the stepwise strategy, however, the inferred cosmological parameters shift toward larger values of $\Omega_m$. Similar high-$\Omega_m$ solutions have also been reported in previous quasar-based cosmological analyses \citep[e.g.][]{Yang:2019vgk}. These shifts are observed for both the full and the low-$z$ QSO subsamples, although the latter yields weaker constraints due to its smaller statistics (see Tables~\ref{tab:lcdm_PPSQuasars}, \ref{tab:wcdm_PPSQuasars}, and \ref{tab:cpl_PPSQuasars}). Within the stepwise approach, where the calibration parameters are fixed prior to cosmological inference, the full QSO sample typically provides tighter constraints simply due to its larger statistical leverage.

It is important to stress that these results reflect a limitation of the stepwise procedure rather than an intrinsic inconsistency of the data. 
By fixing the calibration parameters in advance, potential correlations between the quasar standardization and the cosmological parameters are not propagated into the likelihood, which can lead to biased or unstable estimates when quasars are combined with other probes.

We note that the \textsc{Linmix} calibration parameters inferred from our data are broadly consistent with previous determinations. In particular, for the low-$z$ and high-$z$ QSO subsamples (Table~\ref{tab:linmix2}), the best-fit values of $\gamma$ and $\beta$ agree within $\sim 1$--$2\sigma$ with those reported by \cite{Lusso:2025bhy}. Those authors also report that quasar-based $\Lambda$CDM fits can lead to non-standard values of $\Omega_m$ (see their Table~1), highlighting the sensitivity of quasar cosmological constraints to the calibration/inference strategy.

\subsubsection{Results for $w$CDM}

\begin{table*}
\caption{\label{tab:wcdm} \textbf{Diagnostic stepwise constraints} for the flat $w$CDM parameters obtained in the stepwise analysis using SN+SH0ES, DESI DR2 BAO (DESI), Planck compressed CMB distance priors, and calibrated QSOs (Full QSO Sample and QSO low-$z$, $z<1.43$).}
\begin{tabular}{lccc}
\hline
 &$\Omega_m$ &$w$ & $H_0$\\
\hline
SN+SH0ES & $0.314_{-0.059}^{+0.067} $  & $-0.971^{+0.190}_{-0.120}$ & $73.68\pm 1.000 $  \\ 
SN+SH0ES + DESI & $0.311 \pm 0.008$ & $-1.054 \pm 0.021 $ & $72.90^{+0.980}_{-0.990} $ \\
SN+SH0ES + Planck Compressed & $0.300 \pm 0.003$ & $-1.062  ^{+0.013}_{-0.012}$ & $69.18^{+0.210}_{-0.200} $ \\ 
SN+SH0ES + DESI + QSO low-$z$& $0.579\pm 0.008$ & $-0.696^{+0.018}_{-0.017}$  & - \\ 

SN+SH0ES + DESI + Full QSO Sample & $0.519 \pm 0.007$& $-0.875  ^{+0.020}_{-0.019}$ & - \\ 
SN+SH0ES + Planck Compressed + QSO low-$z$& $0.260 \pm 0.001$ & $-0.892^{+0.007}_{-0.006} $ & - \\ 

SN+SH0ES + Planck Compressed + Full QSO Sample & $0.269\pm 0.002$& $-0.942  ^{+0.008}_{-0.007}$ & - \\ 

\hline
\end{tabular}
\end{table*}
%%%%%%%%

%%%%%%%%

\begin{figure}
   \centering
   \includegraphics[width=3.5 in]{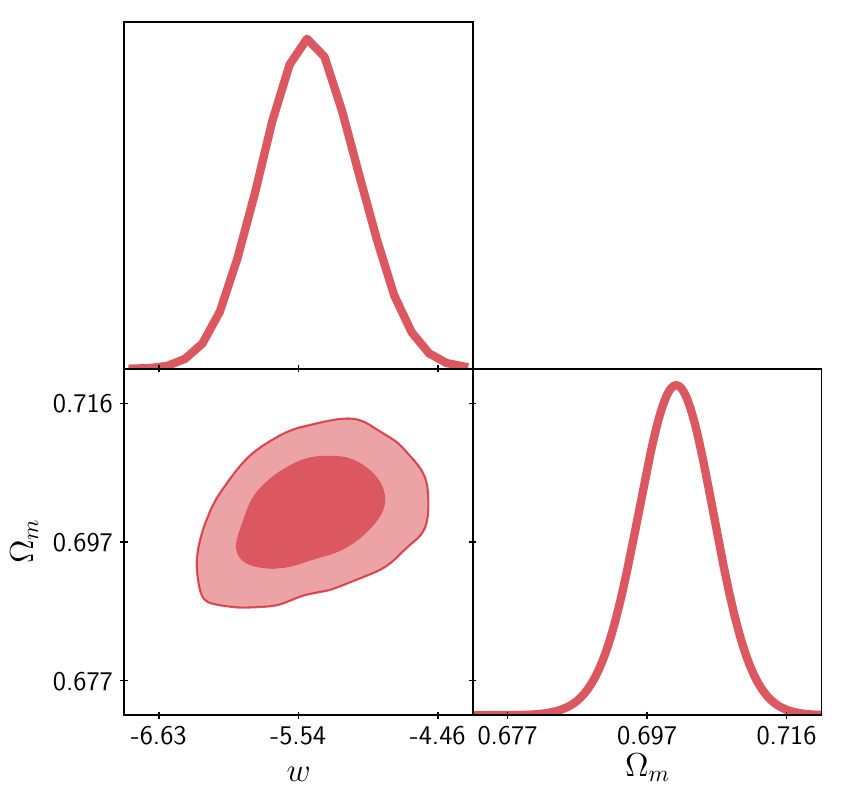}
    \caption{Posterior contours for the $w$CDM model obtained in the diagnostic stepwise analysis using SN+SH0ES combined with the calibrated full QSO sample.}
   \label{wcdm2038}
\end{figure}

\begin{table}
\caption{\label{tab:wcdm_PPSQuasars}
Diagnostic stepwise constraints for the flat $w$CDM model derived using SN+SH0ES in combination with the calibrated QSOs (Full QSO Sample and QSO low-$z$, $z<1.43$).}
\begin{tabular}{lccc}
\hline
 &$\Omega_m$ &$w$ \\
\hline 
SN+SH0ES + QSO low-$z$ & $0.806^{+0.007}_{-0.006}$ & $-5.255^{+0.760}_{-0.660}$  \\ 
SN+SH0ES + Full QSO Sample & $0.701\pm 0.005$ & $-5.454^{+0.360}_{-0.380}$ \\

\hline
\end{tabular}
\end{table}

The corresponding stepwise constraints for the flat $w$CDM model are reported in Table~\ref{tab:wcdm}. As in the $\Lambda$CDM case, the combinations that do not include quasars (e.g. SN+SH0ES, SN+SH0ES+DESI, and SN+SH0ES+Planck Compressed) yield results consistent with standard expectations. When the calibrated QSO sample is added within the stepwise strategy, the inferred parameters shift significantly. In particular, the preferred matter density increases and the equation-of-state parameter departs from its near-$-1$ value, with the full QSO sample providing tighter constraints than the low-$z$ subsample due to its larger statistics. As discussed above, $H_0$ is not constrained by the stepwise QSO-inclusive combinations because the absolute scale is fixed during the calibration stage.

The corresponding constraints from SN+SH0ES combined with the calibrated QSO samples are summarized in Table~\ref{tab:wcdm_PPSQuasars} and Fig.~\ref{wcdm2038} displays the posterior contours for the SN+SH0ES + Full QSO Sample case, while the low-$z$ results are summarized in the table only. In this stepwise configuration, the inferred parameters move to strongly displaced values with respect to standard expectations, with $\Omega_m \gtrsim 0.7$ and very negative $w$. This suggests that, when calibration parameters are fixed prior to cosmological inference, residual calibration--cosmology degeneracies may not be fully propagated and can pull the posterior toward extreme regions of parameter space. We therefore interpret these results as illustrating the limitations of the stepwise procedure and use them mainly as diagnostic benchmarks, motivating the joint calibration--cosmology analysis presented below.

\subsubsection{Results for CPL}

%%%%%%%%%%%%%%%%%%%%%%%%%%%%%%%%%%%%%%%
\begin{table*}
\caption{\label{tab:cpl} Diagnostic stepwise constraints for the flat CPL model parameters obtained in the stepwise analysis using SN+SH0ES, DESI DR2 BAO (DESI), Planck compressed CMB distance priors, and calibrated QSOs (Full QSO Sample and QSO low-$z$, $z<1.43$).}
\begin{tabular}{lcccc}
\hline
 &$\Omega_m$ &$w_{0}$ & $w_{a}$ & $H_0$\\
\hline
SN+SH0ES & $0.329_{-0.052}^{+0.120}$ & $-0.970_{-0.140}^{+0.180}$ & $-0.4974_{-0.660}^{+1.500}$ & $73.68_{-1.000}^{+0.970}$ \\ 
SN+SH0ES + DESI & $0.312 \pm{0.008}$ & $-1.048^{+0.024}_{-0.023}$  & $-0.030^{+0.021}_{-0.070}$ & $72.89^{+0.980}_{-1.000}$  \\
SN+SH0ES + Planck Compressed & $0.300 \pm 0.003$ & $-1.053 ^{+0.020}_{-0.015}$& $-0.036^{+0.018}_{-0.064}$ & $69.18^{+0.200}_{-0.210}$  \\ 
SN+SH0ES + DESI + QSO low-$z$& $0.574\pm 0.008$ & $-0.223^{+0.062}_{-0.065}$  & $-2.395^{+0.370}_{-0.320}$ & - \\ 

SN+SH0ES + DESI + Full QSO Sample & $0.519\pm 0.007$& $-0.882\pm 0.022$ & $0.030^{+0.070}_{-0.021}   $ & - \\ 
SN+SH0ES + Planck Compressed + QSO low-$z$&  $0.258 \pm 0.001$ & $-0.858^{+0.007}_{-0.006}$ & $-0.098^{+0.0003}_{-0.001}$ & - \\ 

SN+SH0ES + Planck Compressed + Full QSO Sample & $0.265 \pm 0.002$ & $-0.327 ^{+0.034}_{-0.035}$ & $-2.306^{+0.150}_{-0.140}$ & - \\ 

\hline
\end{tabular}
\end{table*}

%%%%%%%%%%%%%%%%%%%%%%%%%%%%%
\begin{figure}
   \centering
   \includegraphics[width=3.5 in]{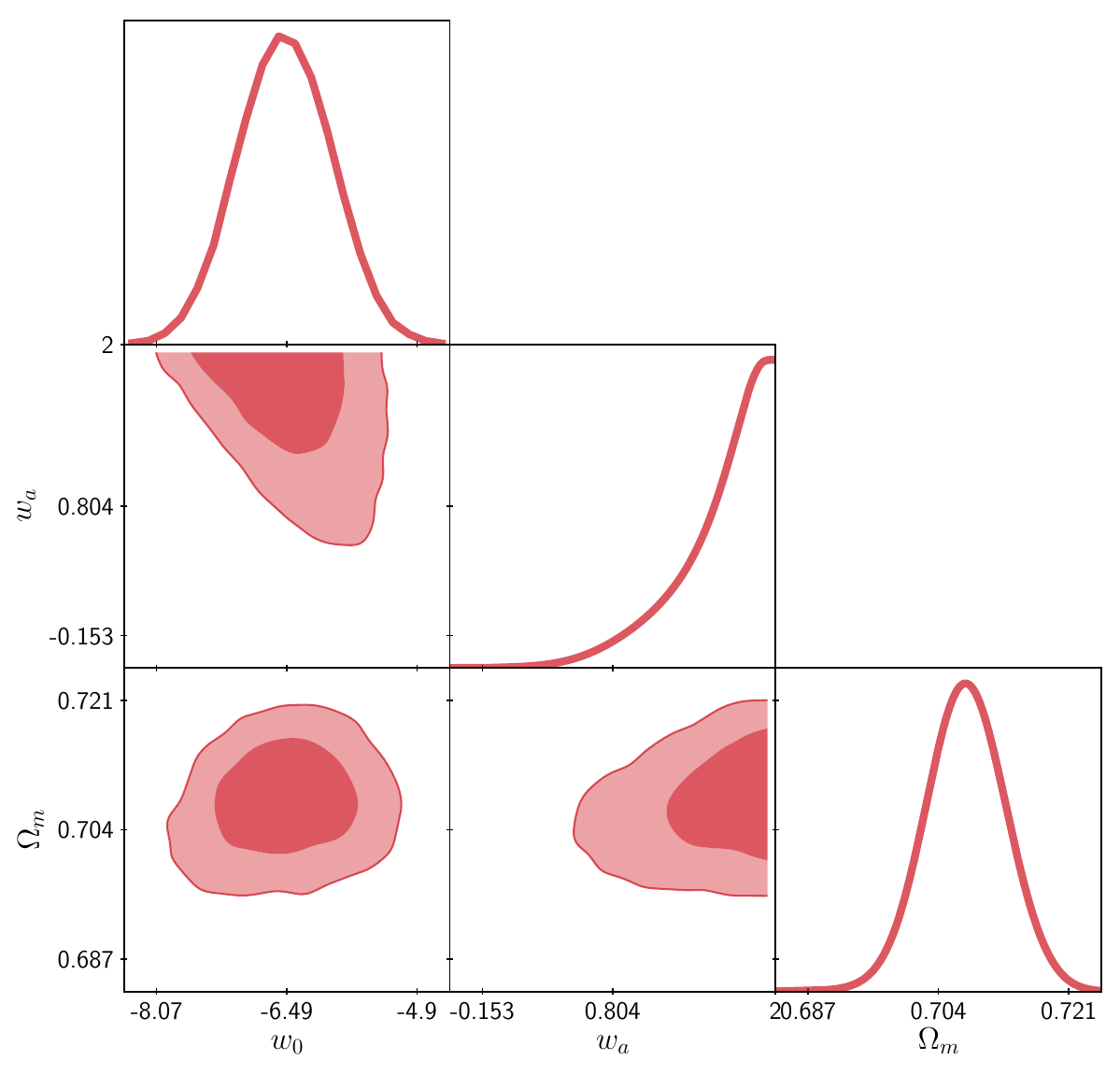}
    \caption{Posterior contours for the CPL model obtained in the diagnostic stepwise analysis using SN+SH0ES combined with the calibrated full QSO sample.}
   \label{cpl2038}
\end{figure}

\begin{table}
\caption{\label{tab:cpl_PPSQuasars}
Diagnostic stepwise constraints for the flat CPL model derived using SN+SH0ES in combination with the calibrated QSOs (Full QSO Sample and QSO low-$z$, $z<1.43$).}
\begin{tabular}{lcccc}
\hline
 &$\Omega_m$ &$w_{0}$ &$w_{a}$ \\
\hline 
SN+SH0ES + QSO low-$z$ & $0.807 ^{+0.008}_{-0.007}$  & $-5.335^{+0.980}_{-0.710}$  & $0.467^{+1.50}_{-0.46}$  \\ 
SN+SH0ES + Full QSO Sample & $0.708^{+0.006}_{-0.005}$ & $-6.506^{+0.560}_{-0.570}$ & $1.603^{+0.400}_{-0.091}$ \\

\hline
\end{tabular}
\end{table}

The stepwise constraints for the CPL model are summarized in Table~\ref{tab:cpl}. For the baseline combinations that do not include quasars (SN+SH0ES, SN+SH0ES+DESI, and SN+SH0ES+Planck Compressed), the inferred parameters remain consistent with the standard cosmological picture, with $w_0 \simeq -1$ and $w_a$ compatible with zero within uncertainties.

When the calibrated QSO samples are incorporated within the stepwise framework, however, the inferred parameters shift substantially. In particular, the preferred matter density increases and the dark-energy parameters $(w_0, w_a)$ move toward extreme values, with the full QSO sample yielding tighter but strongly displaced constraints compared to the low-$z$ subsample. As in the previous models, $H_0$ is not constrained in these combinations because the overall distance scale is fixed during the calibration stage.

We interpret these shifts as a limitation of the stepwise procedure. By fixing the calibration parameters prior to cosmological inference, correlations between the quasar standardization and the cosmological parameters are not fully propagated, which can drive the posterior toward non-standard regions of parameter space. For this reason, the stepwise CPL results should be regarded primarily as diagnostic benchmarks, and a more self-consistent joint calibration--cosmology analysis is presented below.

For visualization, Fig.~\ref{cpl2038} shows the posterior contours for the SN+SH0ES + Full QSO Sample combination, while the low-$z$ results are summarized in the tables only.

\subsection{Joint calibration–cosmology inference}

A potential concern in quasar standard-candle analyses is the covariance between calibration and cosmological parameters. In stepwise strategies, where the calibration is determined first and then treated as fixed, these correlations are not fully propagated and the inference may become internally inconsistent when quasars are combined with other probes.

To address this, we adopt a joint inference approach in which the quasar calibration and cosmological parameters are sampled simultaneously within a single likelihood. This construction avoids the mismatch between the calibration-stage expansion history and the cosmological-stage expansion history that can arise in the stepwise procedure.

For each cosmological scenario ($\Lambda$CDM, $w$CDM, CPL), we therefore sample a parameter vector of the form
$\theta=\{\theta_{\rm cosmo}, \theta_{\rm cal}\}$,
where $\theta_{\rm cal}\equiv\{\beta,\gamma,\delta\}$ and, when supernovae are included, the nuisance parameter $M_{\rm SN}$. The cosmological parameter set is $\{\Omega_m,H_0\}$ for $\Lambda$CDM, $\{\Omega_m,w,H_0\}$ for $w$CDM, and $\{\Omega_m,w_0,w_a,H_0\}$ for CPL.

To quantify the impact of external information and dataset choices within the same joint calibration cosmology framework, we consider three controlled cases:

\begin{itemize}
    \item \textbf{Case A: QSO+SN+SH0ES.} Joint inference using the full QSO and SN samples, including the SH0ES $H_0$ information implemented as a Gaussian likelihood term.
    
    \item \textbf{Case B: SN+SH0ES (reference).} SN-only configuration (no quasars), included as a consistency check against standard SN-based constraints when the SH0ES $H_0$ term is included.
    
    \item \textbf{Case C: QSO+SN (no SH0ES).} Joint inference using QSO+SN, without any external $H_0$ information.
\end{itemize}

In the figures below, we compare Cases A and C to isolate the impact of the SH0ES $H_0$ term within the same QSO+SN joint analysis. Full marginalized constraints for Cases A/B/C are reported in Tables~\ref{tab:lcdm_cases}--\ref{tab:cpl_cases_ABC}.

\subsubsection{Results for $\Lambda$CDM}

\begin{figure}
   \centering
   \includegraphics[width=3.3 in]{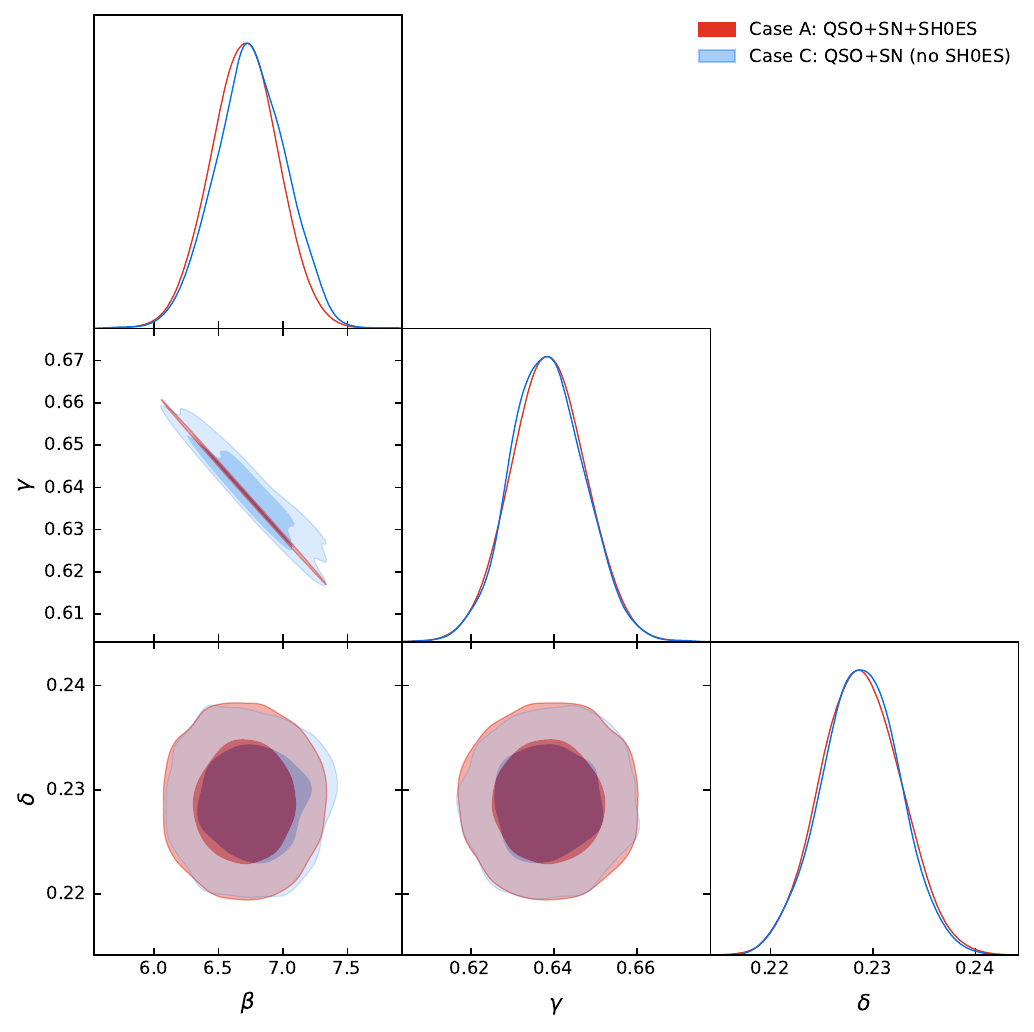}
    \caption{Quasar-calibration posteriors $(\gamma,\beta,\delta)$ from the joint $\Lambda$CDM analysis, shown for Case A (QSO+SN+SH0ES) and Case C (QSO+SN, no SH0ES). Contours indicate the 68\% and 95\% credible regions.}
   \label{fig:cases_calibration_lcdm}
\end{figure}

\begin{figure}
   \centering
   \includegraphics[width=3.3 in]{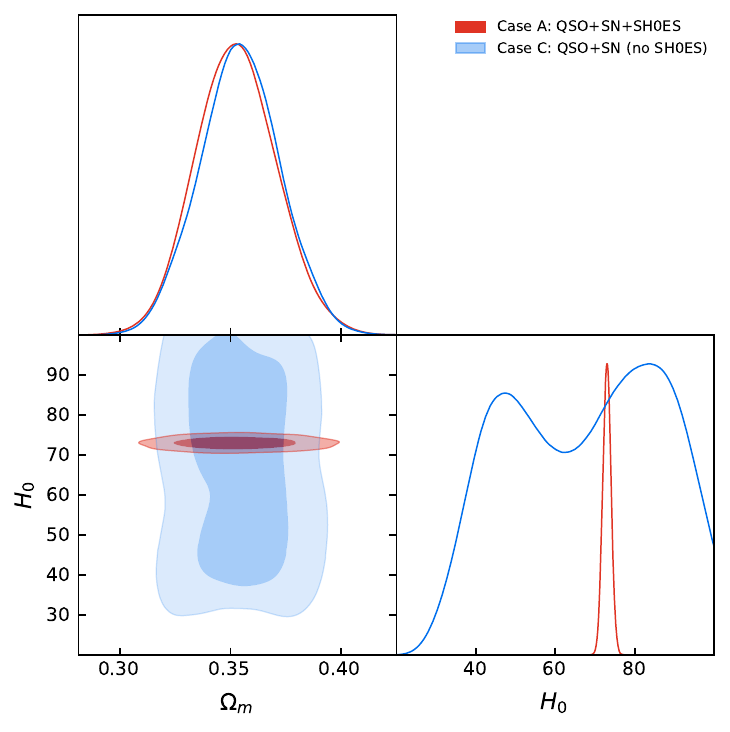}
    \caption{$\Lambda$CDM posterior constraints for Cases A (QSO+SN+SH0ES) and C (QSO+SN, no SH0ES) in the $(\Omega_m, H_0)$ plane. Filled contours show the 68\% and 95\% credible regions. The comparison highlights the impact of the SH0ES $H_0$ constraint (implemented as a Gaussian likelihood term), which primarily tightens the $H_0$ posterior while leaving $\Omega_m$ nearly unchanged.}
   \label{fig:cases_cosmo_lcdm}
\end{figure}

Because all parameters are sampled jointly, calibration uncertainties are consistently propagated into the cosmological constraints. Figure~\ref{fig:cases_calibration_lcdm} shows the posteriors for the quasar-calibration parameters $(\gamma,\beta,\delta)$ in Cases A and C, while Fig.~\ref{fig:cases_cosmo_lcdm} shows the corresponding cosmological constraints in the $(\Omega_m,H_0)$ plane. Table~\ref{tab:lcdm_cases} summarizes the marginalized results for Cases A, B, and C.

We find that $\Omega_m$ is consistent between Cases A and C
(Fig.~\ref{fig:cases_cosmo_lcdm}), indicating that it is primarily
constrained by the QSO+SN combination. In contrast, removing the SH0ES
term mainly affects the absolute calibration: without the external
absolute-distance anchor, the $H_0$ posterior becomes broad and
non-Gaussian due to its strong degeneracy with $M_{\rm SN}$. For this
reason, in Case C we report central 68\% credible intervals for $H_0$
and $M_{\rm SN}$, rather than point estimates with asymmetric errors.
No appreciable shift in $\Omega_m$ is observed.

This contrasts with the stepwise analysis, where fixing the calibration parameters can lead to displaced or unstable solutions once quasars are included. The joint approach therefore provides a more self-consistent inference for QSO-based cosmological constraints.

%%%%%%%%%%%%%%%%%%%%%%%
\begin{table*}
\centering
\caption{$\Lambda$CDM constraints for the three analysis cases. In Cases A and C the QSO calibration parameters are sampled jointly with cosmology; Case B (SN+SH0ES: reference) does not include QSO calibration parameters. For approximately Gaussian posteriors we quote the median and the 68\% credible interval. For the non-Gaussian Case C posteriors of $H_0$ and $M_{\rm SN}$, we report the central 68\% credible interval. For the whole analysis we have used the full QSO sample.}
\label{tab:lcdm_cases}
\begin{tabular}{lccc}
\hline
Parameter & Case A: QSO+SN+SH0ES & Case B: SN+SH0ES & Case C: QSO+SN (no SH0ES) \\
\hline
$\Omega_m$ 
& $0.353^{+0.018}_{-0.018}$ 
& $0.334^{+0.019}_{-0.018}$ 
& $0.354^{+0.018}_{-0.019}$ \\

$H_0$ [km s$^{-1}$ Mpc$^{-1}$] 
& $72.998^{+1.038}_{-1.013}$ 
& $73.030^{+1.029}_{-1.020}$ 
& $[42.5,\,88.0]$ \\

$\beta$ 
& $6.701^{+0.261}_{-0.264}$ 
& -- 
& $6.749^{+0.276}_{-0.276}$ \\

$\gamma$ 
& $0.639^{+0.009}_{-0.009}$ 
& -- 
& $0.639^{+0.009}_{-0.009}$ \\

$\delta$ 
& $0.229^{+0.004}_{-0.004}$ 
& -- 
& $0.229^{+0.004}_{-0.004}$ \\

$M_{\rm SN}$ 
& $-19.253^{+0.031}_{-0.031}$ 
& $-19.258^{+0.031}_{-0.031}$ 
& $[-20.43,\,-18.85]$ \\
\hline
\end{tabular}
\end{table*}

%%%%%%%%%%%%%%%%%%%%%%%

%%%%%%%%

%%%%%%%

\subsubsection{Results for $w$CDM}

\begin{figure}
   \centering

   \includegraphics[width=3.3 in]{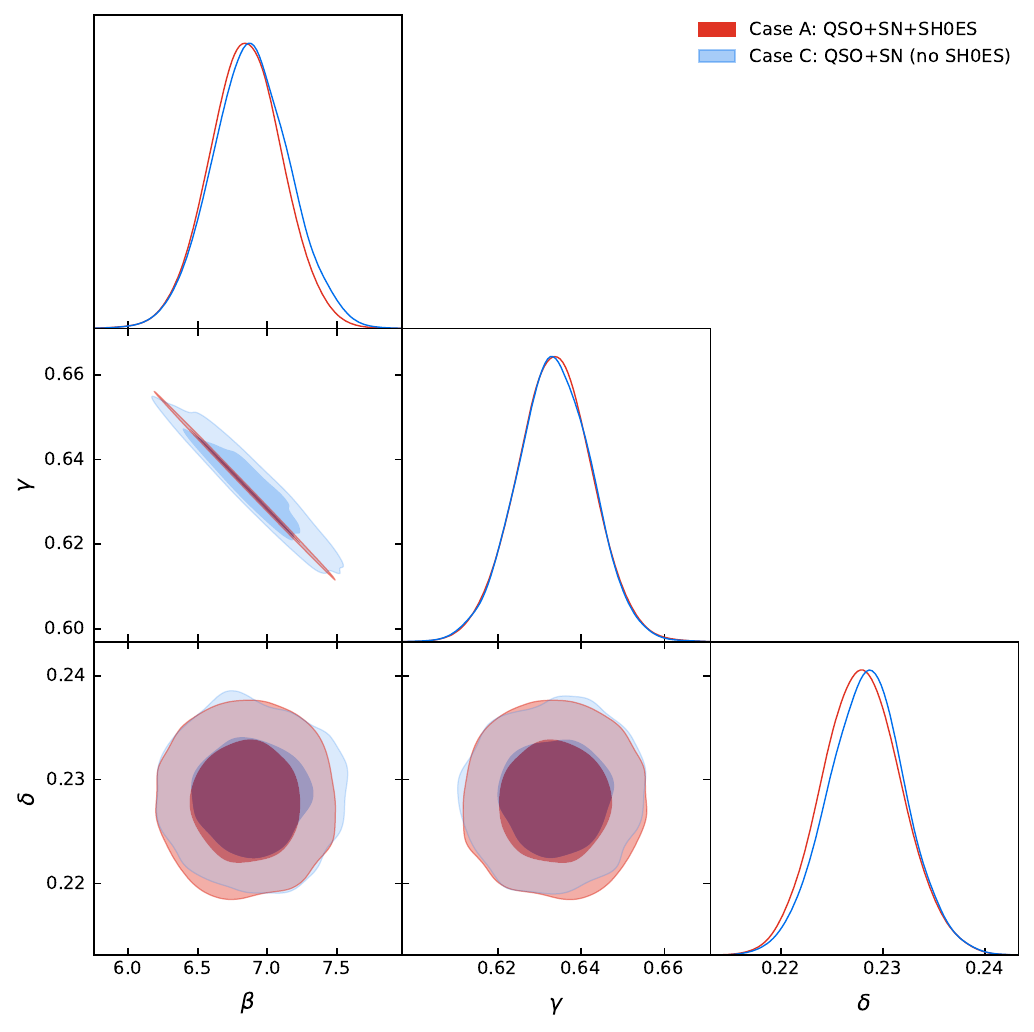}
    \caption{Quasar-calibration posteriors $(\gamma,\beta,\delta)$ from the joint $w$CDM analysis, shown for Case A (QSO+SN+SH0ES) and Case C (QSO+SN, no SH0ES). Contours indicate the 68\% and 95\% credible regions.}
   \label{fig:cases_calibration_wcdm}
\end{figure}

\begin{figure}
   \centering
   \includegraphics[width=3.3 in]{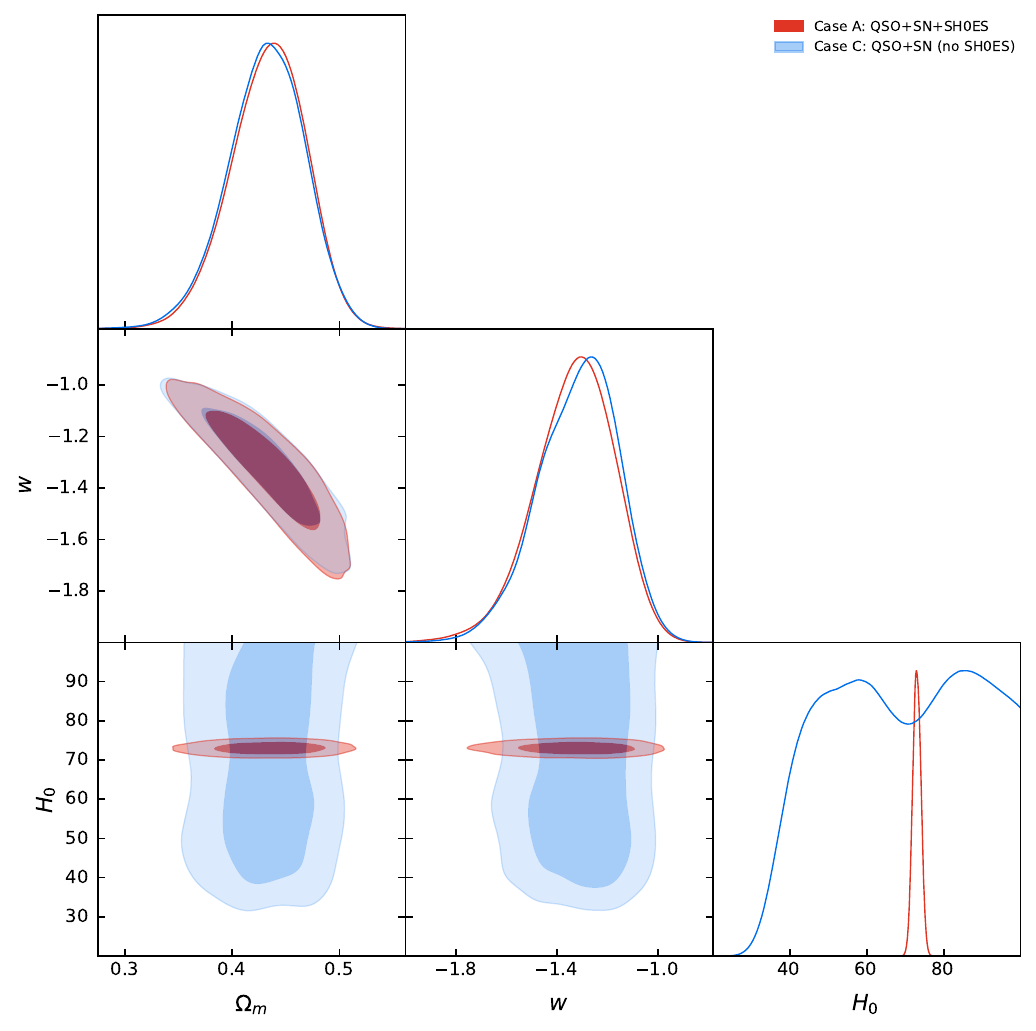}
    \caption{$w$CDM posterior constraints for Cases A (QSO+SN+SH0ES) and C (QSO+SN, no SH0ES), shown as a triangle plot for $(\Omega_m, w, H_0)$. Contours indicate the 68\% and 95\% credible regions. The comparison illustrates the impact of including the SH0ES $H_0$ likelihood term within the same joint QSO+SN analysis.}
   \label{fig:cases_cosmo_wcdm}
\end{figure}

The same joint strategy is applied to the $w$CDM model. Figures~\ref{fig:cases_calibration_wcdm} and \ref{fig:cases_cosmo_wcdm}, together with Table~\ref{tab:wcdm_cases_ABC}, show that the cosmological constraints remain consistent between Cases A and C.

Allowing $w\neq -1$ introduces additional degeneracies involving $H_0$ and the calibration parameters, which are naturally captured by the joint sampling. When the SH0ES term is removed (Case C), the constraints on $H_0$ and $M_{\rm SN}$ broaden, as expected. The posteriors for $\Omega_m$ and $w$ remain consistent within uncertainties.

Overall, the joint analysis yields stable parameter constraints and avoids the extreme solutions encountered in the stepwise configuration.

\begin{table*}
\centering
\caption{wCDM constraints for the three analysis cases. In Cases A and C the QSO calibration parameters are sampled jointly with cosmology; Case B (SN+SH0ES: reference) does not include QSO calibration parameters. For approximately Gaussian posteriors we quote the median and the 68\% credible interval. For the non-Gaussian Case C posteriors of $H_0$ and $M_{\rm SN}$, we report the central 68\% credible interval. For the whole analysis we have used the full QSO sample.}
\label{tab:wcdm_cases_ABC}
\begin{tabular}{lccc}
\hline\hline
Parameter & Case A (QSO+SN+SH0ES) & Case B (SN+SH0ES) & Case C (QSO+SN; no SH0ES) \\
\hline
$\Omega_m$          
& $0.435^{+0.034}_{-0.037}$ 
& $0.291^{+0.064}_{-0.077}$ 
& $0.435^{+0.035}_{-0.037}$ \\

$w$                 
& $-1.321^{+0.151}_{-0.169}$ 
& $-0.893^{+0.141}_{-0.156}$ 
& $-1.322^{+0.152}_{-0.171}$ \\

$H_0$ [km s$^{-1}$ Mpc$^{-1}$]               
& $73.039^{+1.043}_{-1.034}$ 
& $73.038^{+1.030}_{-1.035}$ 
& $[48.31,\,89.65]$ \\

$\beta$             
& $6.839^{+0.268}_{-0.265}$ 
& ---                        
& $6.872^{+0.290}_{-0.280}$ \\

$\gamma$            
& $0.634^{+0.009}_{-0.009}$ 
& ---                        
& $0.634^{+0.009}_{-0.009}$ \\

$\delta$  
& $0.228^{+0.004}_{-0.004}$ 
& ---                        
& $0.228^{+0.004}_{-0.004}$ \\

$M_{\rm SN}$        
& $-19.264^{+0.032}_{-0.032}$ 
& $-19.254^{+0.032}_{-0.032}$ 
& $[-20.16,\,-18.82]$ \\
\hline\hline
\end{tabular}
\end{table*}

%%%%%%%%

%%%%%%%

\subsubsection{Results for CPL}

\begin{figure}
   \centering
   \includegraphics[width=3.3 in]{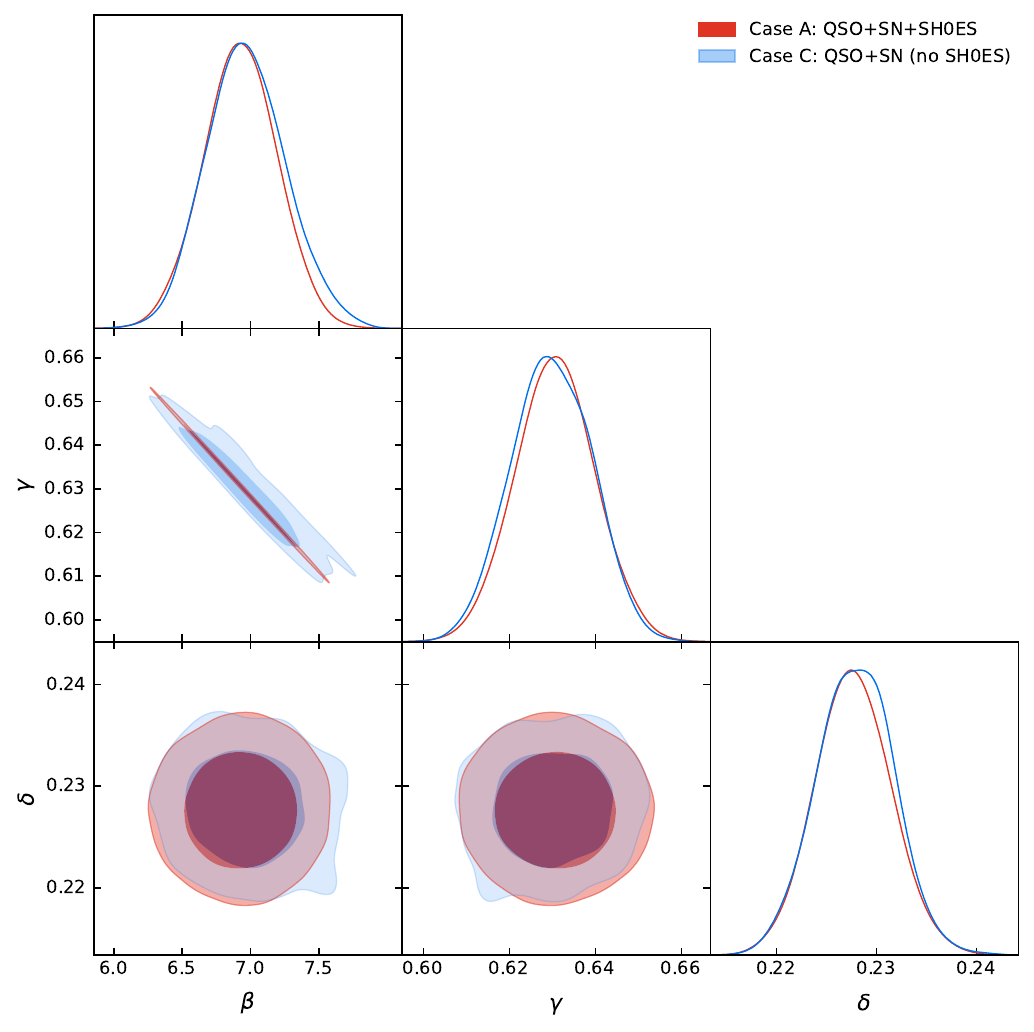}
    \caption{Quasar-calibration posteriors $(\gamma,\beta,\delta)$ from the joint CPL fit, shown for Case A (QSO+SN+SH0ES) and Case C (QSO+SN, no SH0ES). Contours indicate the 68\% and 95\% credible regions.}
   \label{fig:cases_calibration_cpl}
\end{figure}

\begin{figure*}
   \centering
   \includegraphics[width=5.5 in]{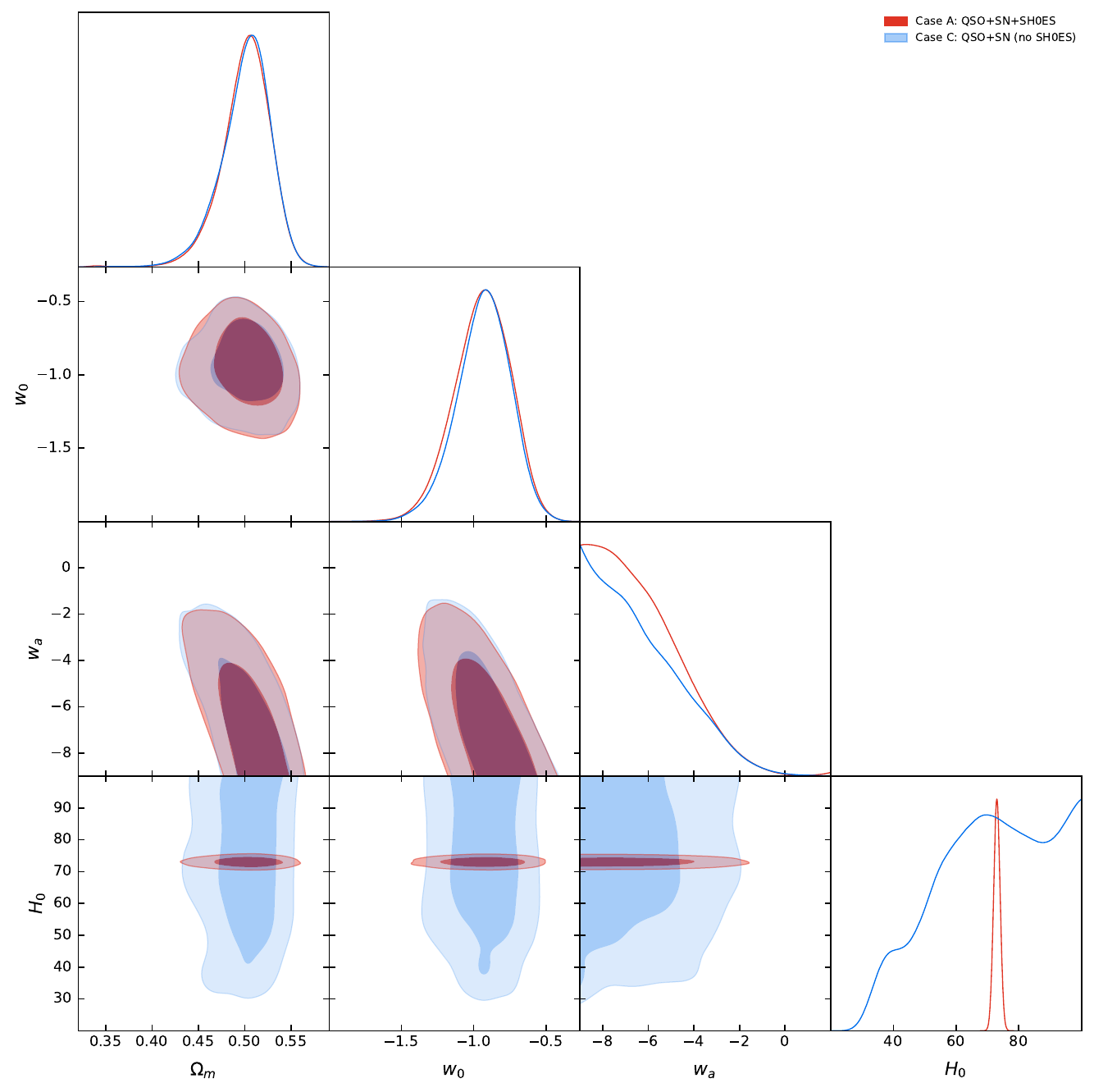}
    \caption{CPL posterior constraints for Cases A (QSO+SN+SH0ES) and C (QSO+SN, no SH0ES), shown as a triangle plot for $(\Omega_m, w_0, w_a, H_0)$. Contours indicate the 68\% and 95\% credible regions. Including the SH0ES-based Gaussian prior in Case A primarily tightens the $H_0$ constraint and modifies its degeneracies with $(w_0, w_a)$.}

   \label{fig:cases_cosmo_cpl}
\end{figure*}

For the CPL model we sample jointly $(\Omega_m,w_0,w_a,H_0)$ together with $(\beta,\gamma,\delta)$ (and $M_{\rm SN}$ when SNe are included). Figure~\ref{fig:cases_calibration_cpl} shows the calibration posteriors in Cases A and C, while Fig.~\ref{fig:cases_cosmo_cpl} shows the corresponding cosmological posteriors. The marginalized constraints for Cases A, B, and C are reported in Table~\ref{tab:cpl_cases_ABC}.

Despite the larger parameter space and the degeneracies between $(w_0,w_a)$ and $H_0$, the joint inference remains stable. As in the previous models, removing the SH0ES term primarily enlarges the uncertainty on $H_0$ (Fig.~\ref{fig:cases_cosmo_cpl}), while the preferred region of $(\Omega_m,w_0,w_a)$ remains broadly consistent. The calibration posteriors are also consistent between Cases A and C (Fig.~\ref{fig:cases_calibration_cpl}), indicating that the QSO standardization is not strongly driven by the external $H_0$ information.

These results show that simultaneous calibration–cosmology inference yields self-consistent constraints in extended dark-energy models and avoids the unstable parameter values encountered in the stepwise analysis.

\begin{table*}
\centering
\caption{CPL constraints for the three analysis cases. In Cases A and C the QSO calibration parameters are sampled jointly with cosmology; Case B (SN+SH0ES: reference) does not include QSO calibration parameters. For approximately Gaussian posteriors we quote the median and the 68\% credible interval. For broad, non-Gaussian, or prior-affected posteriors, such as $w_a$ and the Case C posteriors of $H_0$ and $M_{\rm SN}$, square brackets indicate central 68\% marginalized credible intervals rather than the full posterior support. For the whole analysis we have used the full QSO sample.}

\label{tab:cpl_cases_ABC}
\begin{tabular}{lccc}
\hline\hline
Parameter & Case A (QSO+SN+SH0ES) & Case B (SN+SH0ES) & Case C (QSO+SN; no SH0ES) \\
\hline
$\Omega_m$          
& $0.504^{+0.023}_{-0.027}$ 
& $0.374^{+0.084}_{-0.147}$ 
& $0.504^{+0.023}_{-0.028}$ \\

$w_0$               
& $-0.926^{+0.183}_{-0.202}$ 
& $-0.889^{+0.147}_{-0.158}$ 
& $-0.932^{+0.180}_{-0.203}$ \\

$w_a$               
& $[-8.30,\,-4.40]$ 
& $[-4.11,\,0.48]$ 
& $[-8.27,\,-4.28]$ \\

$H_0$ [km s$^{-1}$ Mpc$^{-1}$]               
& $73.057^{+1.029}_{-1.049}$ 
& $73.041^{+1.033}_{-1.045}$ 
& $[47.63,\,91.12]$ \\

$\beta$             
& $6.936^{+0.269}_{-0.269}$  
& ---                        
& $6.971^{+0.292}_{-0.287}$ \\

$\gamma$            
& $0.630^{+0.009}_{-0.009}$  
& ---                        
& $0.630^{+0.009}_{-0.009}$ \\

$\delta$  
& $0.228^{+0.004}_{-0.004}$  
& ---                        
& $0.228^{+0.004}_{-0.004}$ \\

$M_{\rm SN}$        
& $-19.245^{+0.032}_{-0.033}$ 
& $-19.251^{+0.032}_{-0.033}$ 
& $[-20.17,\,-18.77]$ \\
\hline\hline
\end{tabular}
\end{table*}

%%%%%%%%

%%%%%%%

\subsubsection{Sensitivity to QSO redshift coverage}

\begin{figure}
   \centering
   \includegraphics[width=3.3 in]{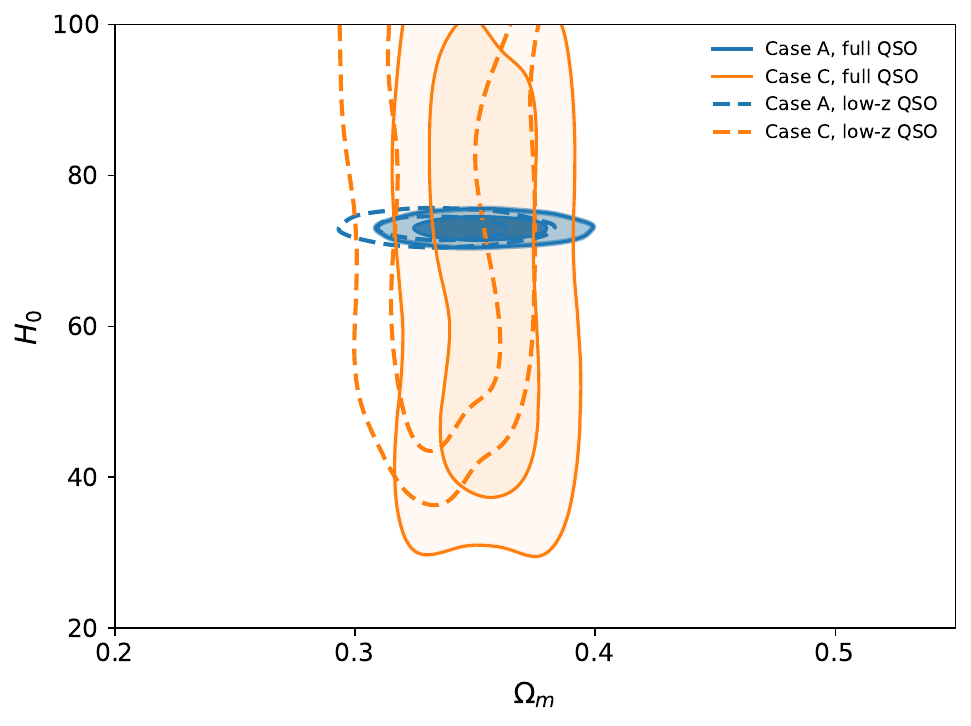}
    \caption{$\Lambda$CDM posterior constraints in the $(\Omega_m,H_0)$ plane for Cases A and C, corresponding to QSO+SN with and without the SH0ES $H_0$ information, respectively. Solid filled contours denote the full QSO sample, while dashed contours denote the low-$z$ subsample. Contours indicate the 68\% and 95\% credible regions.}
   \label{fig:cases_lcdm_full-low}
\end{figure}

\begin{figure}
   \centering
   \includegraphics[width=3.3 in]{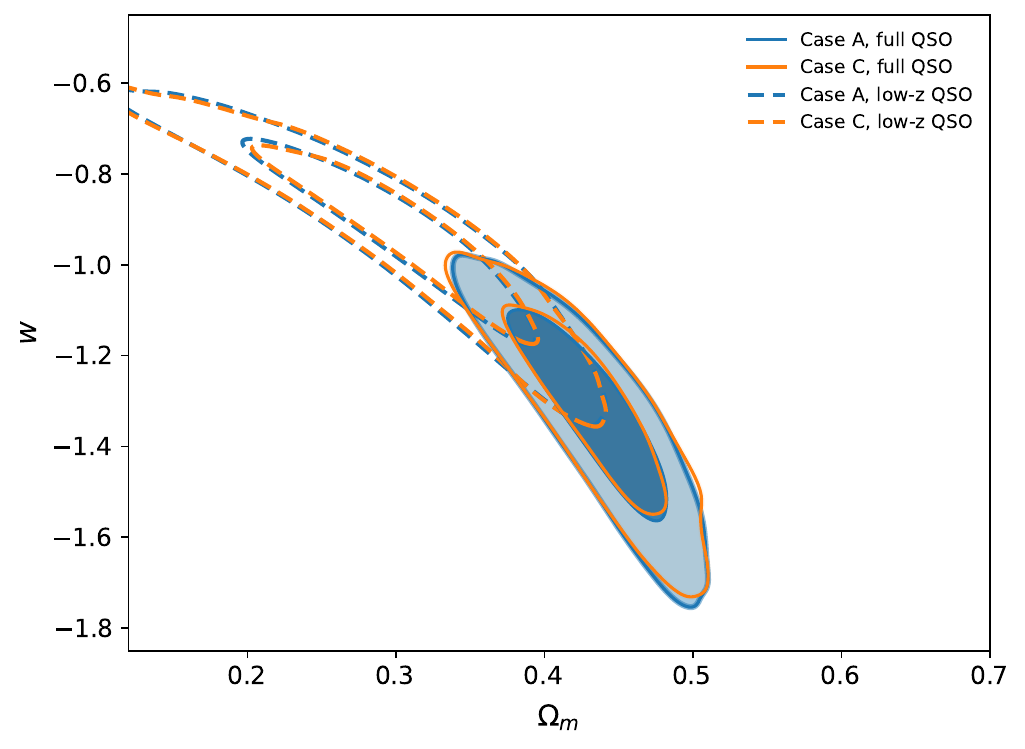}
    \caption{$w$CDM posterior constraints in the $(\Omega_m, w)$ plane for Cases A and C, corresponding to QSO+SN with and without the SH0ES $H_0$ information, respectively. Solid filled contours denote the full QSO sample, while dashed contours denote the low-$z$ subsample. Contours indicate the 68\% and 95\% credible regions.}
   \label{fig:cases_wcdm_full-low}
\end{figure}

\begin{figure}
   \centering
   \includegraphics[width=3.3 in]{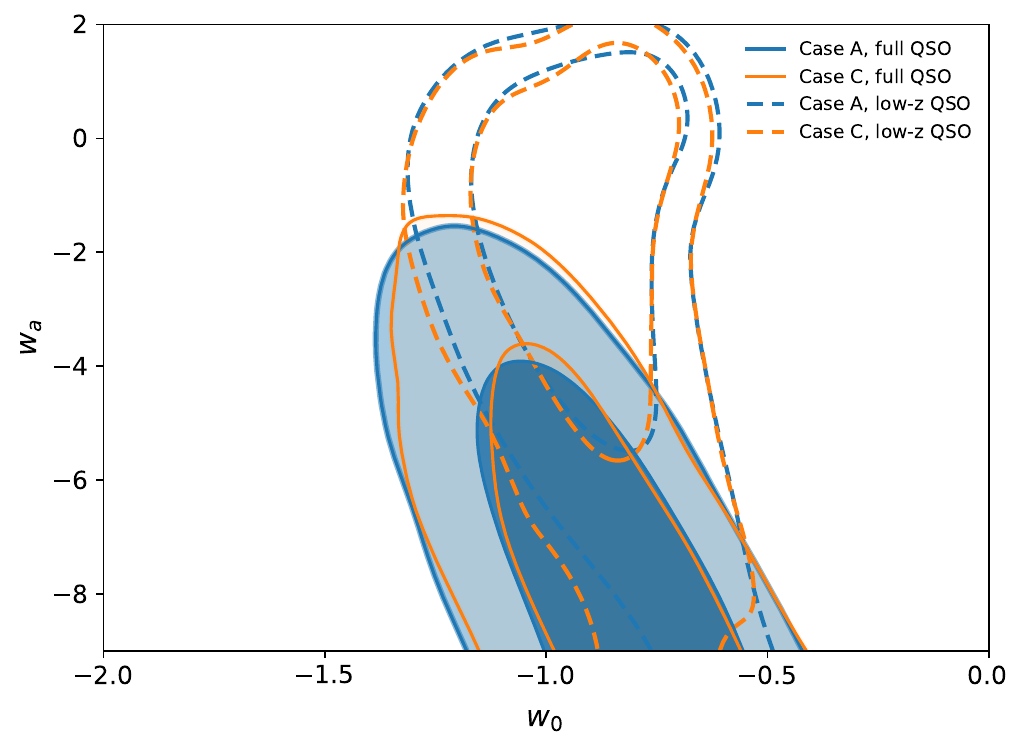}
    \caption{CPL posterior constraints in the $(w_0, w_a)$ plane for Cases A and C, corresponding to QSO+SN with and without the SH0ES $H_0$ information, respectively. Solid filled contours denote the full QSO sample, while dashed contours denote the low-$z$ subsample. Contours indicate the 68\% and 95\% credible regions.}
   \label{fig:cases_cpl_full-low}
\end{figure}

\begin{figure*}
    \centering
    \begin{subfigure}[t]{0.4\textwidth}
        \centering
        \includegraphics[width=\linewidth]{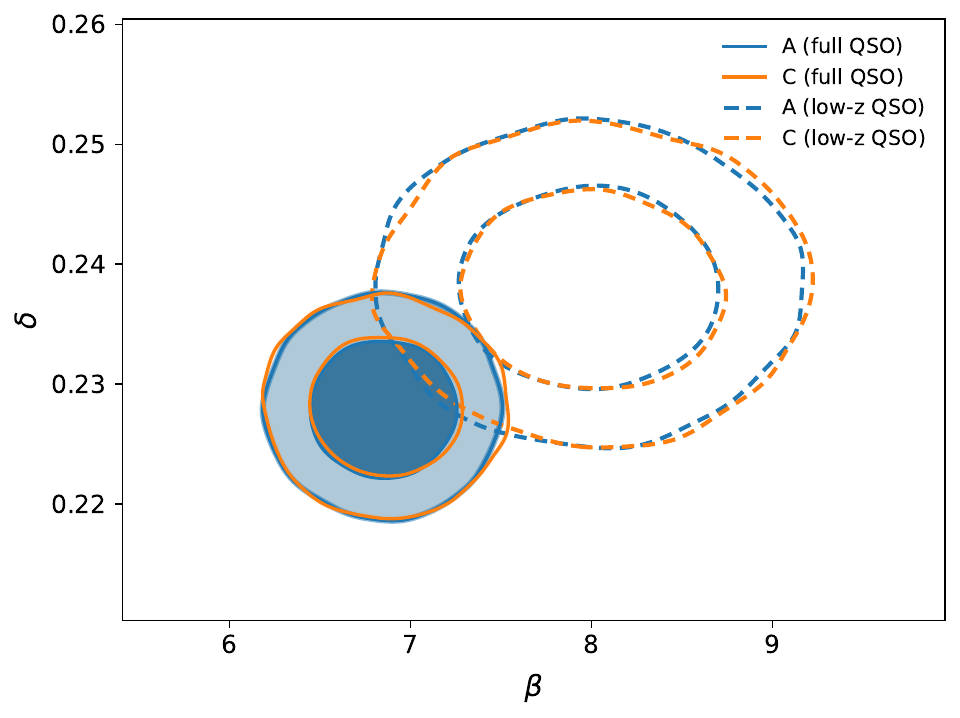}
        \caption{$(\beta, \delta)$ plane.}
        \label{fig:wcdm_calib_delta_gamma}
    \end{subfigure}\hspace{6mm}
    \begin{subfigure}[t]{0.4\textwidth}
        \centering
        \includegraphics[width=\linewidth]{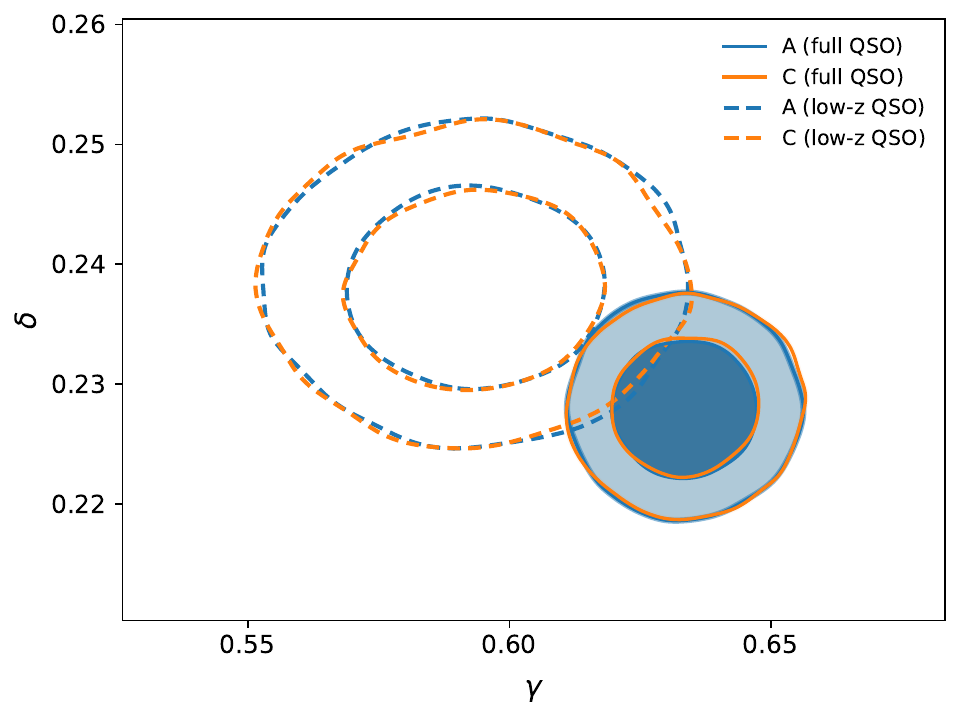}
        \caption{$(\gamma,\delta)$ plane.}
        \label{fig:wcdm_calib_delta_beta}
    \end{subfigure}

    \caption{Quasar-calibration constraints from the joint $w$CDM analysis. Filled contours correspond to the full-$z$ quasar sample, while dashed contours show the low-$z$ subsample. In both panels, Case A (QSO+SN+SH0ES) and Case C (QSO+SN, no SH0ES) are overlaid; contours indicate the 68\% and 95\% credible regions.}
    \label{fig:wcdm_calibration_full_vs_low}
\end{figure*}

In addition to the full-redshift QSO sample, we repeat the joint analysis using a low-$z$ subsample ($z<1.43$). This choice is motivated by previous studies suggesting that quasar distances inferred from the $L_X$--$L_{\rm UV}$ relation are more stable and in closer agreement with standard $\Lambda$CDM expectations up to $z\sim 1.5$, while potential selection or redshift-dependent effects may become more relevant at higher redshift \citep{Lusso:2025bhy}. The low-$z$ selection is therefore often adopted as a conservative subsample that may be less sensitive to redshift-dependent effects, although at the cost of reduced statistics and hence broader posteriors. For completeness, the full marginalized constraints obtained with the low-$z$ subsample for all models and analysis cases are reported in Appendix~A.

Figures~\ref{fig:cases_lcdm_full-low}--\ref{fig:cases_cpl_full-low} compare the cosmological constraints obtained with the two QSO selections within the same joint framework, for Cases A (with the SH0ES $H_0$ prior) and C (no SH0ES). As expected from its reduced statistics, the low-$z$ subsample generally yields broader posteriors. In addition, we observe systematic shifts in the preferred parameter regions relative to the full-sample results. These shifts are moderate for $\Lambda$CDM, and become more pronounced in the extended models ($w$CDM and CPL), where additional degeneracies enhance the impact of calibration differences.

This behavior indicates that the inferred cosmological constraints retain some sensitivity to the QSO redshift coverage through the calibration parameters. Within our joint framework, such effects are consistently propagated into the posterior distributions rather than being artificially fixed, allowing us to directly quantify their impact.

Importantly, the calibration parameters show noticeable differences between the full-$z$ and low-$z$ selections. 

Figure~\ref{fig:wcdm_calibration_full_vs_low} illustrates this effect for the joint $w$CDM analysis, where changes in $(\beta,\gamma,\delta)$ partially track the differences in the cosmological posteriors. This supports the interpretation that part of the sensitivity to QSO redshift coverage arises from sample-dependent calibration effects, which are consistently incorporated into the cosmological inference in the joint approach.

Overall, the results presented in this section reinforce the importance of treating quasar calibration and cosmological inference in a self-consistent way. While the stepwise approach remains useful as a diagnostic benchmark, the joint analysis explicitly captures the covariance between calibration and cosmological parameters and yields stable constraints for all the dark-energy scenarios explored here. We therefore adopt the joint constraints as our primary results, and use the low-$z$ QSO subsample as a complementary robustness test against possible redshift-dependent effects.

%%%%%%%%%%%%%%%%%%%%%%%%%%%%%%%%%%%%%%%%%%%%%%%%%%%%%%%%%%%%%%%%%%%%%%%%%%%%%%%%

\section{Conclusions}
\label{sec:conclusions}

A central challenge in the use of quasars as standard candles is the circularity problem: the luminosity--distance relation depends on cosmology, yet quasars are simultaneously employed to constrain that same cosmology. In this work, we addressed this issue by adopting a model-independent calibration strategy based on cosmic-chronometer $H(z)$ measurements up to $z=1.43$, following the selection criteria of \cite{Lusso:2020pdb} and retaining 2038 quasars from the original sample.

Specifically, we reconstruct the expansion history at low redshift using $H(z)$ data and use it to determine luminosity distances independently of any dark-energy parameterization. This procedure provides an external distance scale against which the non-linear $L_X$--$L_{\rm UV}$ relation can be calibrated. In this sense, our approach is analogous to the anchoring of Type Ia supernovae, where an independent local distance indicator is used to fix the absolute magnitude prior to cosmological fits. Here, the $H(z)$ reconstruction plays a similar anchoring role, allowing the quasar luminosity scale to be established without assuming a fiducial background cosmology at the calibration stage.

We first explored a traditional stepwise strategy, in which the calibration parameters are determined independently and subsequently fixed when constraining $\Lambda$CDM, $w$CDM and CPL models. While this approach provides a useful baseline and facilitates comparison with previous analyses, it has an intrinsic limitation: the expansion history used to calibrate the quasars is fixed by the CC-based reconstruction and is not necessarily the same as the one preferred by the cosmological model in the second stage. Moreover, once the normalization parameter of the quasar relation is fixed, the absolute distance scale of the quasar Hubble diagram is also fixed. Since this scale is degenerate with $H_0$, the stepwise approach cannot fully readjust the quasar calibration and the cosmological parameters simultaneously. These effects can lead to shifted or unstable cosmological constraints when quasars are combined with other probes. For this reason, we regard the stepwise results as diagnostic benchmarks rather than as our main cosmological constraints.

To overcome this limitation, we performed a joint calibration--cosmology inference, sampling the quasar-calibration parameters $(\beta,\gamma,\delta)$ simultaneously with the cosmological parameters within a single likelihood. In this framework, calibration uncertainties are consistently marginalized over, and their covariance with cosmology is naturally captured. The resulting constraints are stable and physically reasonable across $\Lambda$CDM, $w$CDM and CPL, although the CPL parameter $w_a$ remains weakly constrained and partially prior-affected. In particular, $\Omega_m$ remains robust when the external SH0ES $H_0$ prior is removed, while the main impact of the prior is to tighten the absolute distance scale. We therefore adopt the joint constraints as our primary results.

We further tested the sensitivity of the inference to QSO redshift coverage by repeating the joint analysis using a low-$z$ subsample ($z<1.43$), motivated by previous studies suggesting that the $L_X$--$L_{\rm UV}$ relation remains relatively stable up to $z\sim1.5$ \citep{Lusso:2025bhy}. The low-$z$ selection, often adopted as a conservative subsample, yields broader posteriors due to reduced statistics and exhibits moderate shifts relative to the full-sample results. Within the joint framework, these differences can be traced to sample-dependent calibration variations that are consistently propagated into the cosmological posteriors.

It is instructive to place these findings in the context of recent literature. Using quasar Hubble diagrams, \cite{Colgain:2022nlb} reported that quasars can appear consistent with Planck--$\Lambda$CDM at low redshift (with $\Omega_m \approx 0.3$) over a restricted range, while also noting that the inferred cosmological parameters vary with the effective redshift of the sample. Subsequently, \cite{Colgain:2024clf} argued that no statistically significant departure from Planck--$\Lambda$CDM is detected at low redshift, and that apparent agreement may depend on restricting the analysis to a limited redshift range and on the construction of the quasar distances. Our results are consistent with the view that cosmological inferences from quasars retain some sensitivity to redshift coverage, and that this sensitivity is closely linked to calibration effects.

At the regression level, the best-fit values of $\gamma$ and $\beta$ obtained from the low-$z$ and high-$z$ samples are broadly consistent with previous studies, supporting the interpretation of the non-linear $L_X$--$L_{\rm UV}$ relation as a genuine physical property of accreting black holes. However, the current level of intrinsic dispersion still limits the constraining power of quasars as precision cosmological probes. Achieving a smaller scatter, for instance at the $\sim 0.2$ dex level reported in carefully selected subsamples \citep{Lusso_2016,Lusso2017AA,Lusso:2020pdb,Signorini:2023nyd}, would significantly enhance their cosmological impact.

Our conclusions are, therefore, twofold. First, quasars can be incorporated into cosmological analyses in a statistically self-consistent manner through joint calibration--cosmology inference, avoiding biases associated with stepwise approaches. Second, while the present data do not yet provide constraints competitive with the most precise low-redshift probes, they offer a promising avenue for extending the cosmic distance ladder to $z>2$, provided that calibration systematics and intrinsic dispersion are further reduced.

Future progress will likely depend on improving the quality and homogeneity of X-ray measurements and on achieving a better understanding of possible redshift-dependent or selection effects in the $L_X$--$L_{\rm UV}$ relation. Methodologically, alternative reconstruction techniques (e.g., splines or Gaussian processes) may serve as robustness checks, and validation with mock catalogs will be important to quantify potential residual biases. Together, these developments will clarify the ultimate potential of quasars as high-redshift standard candles.

\section*{Acknowledgements}

We thank the referee for their careful and constructive comments, which significantly improved the clarity and overall quality of this work.

The work of AM has been sponsored by Conahcyt-Mexico through the Posdoc Project I1200/311/2023. SSN acknowledge financial support by PAPIIT IA101825. JCH and SSN acknowledge financial support by SECIHTI grant CBF 2023-2024-162. JCH acknowledges support from DGAPA-PAPIIT-UNAM grant No. IN110325 "Estudios en cosmología inflacionaria, agujeros negros primordiales y energía oscura" and grant PAPIIT UNAM No. IN114626 “Estudios de ANPs y Contribuciones In-Kind a la colaboración LSST-México". 

\section*{Data availability}

The quasar dataset employed in this study corresponds to the compilation presented by \citet{Lusso:2020pdb}, which is publicly accessible at \url{https://cdsarc.u-strasbg.fr/viz-bin/cat/J/A+A/642/A150}. The data underlying this article will be shared upon reasonable request to the corresponding author.

%%%%%%%%%%%%%%%%%%%%%%%%%%%%%%%%%%%%%%%%%%%%%%%%%%



\bibliographystyle{mnras}
\bibliography{biblio} % if your bibtex file is called example.bib

%%%%%%%%%%%%%%%%%%%%%%%%%%%%%%%%%%%%%%%%%%%%%%%%%%

%%%%%%%%%%%%%%%%% APPENDICES %%%%%%%%%%%%%%%%%%%%%

\appendix

\section{Robustness Tests with the Low-$z$ QSO Subsample}

In this Appendix we report the full joint calibration--cosmology constraints obtained using the low-$z$ QSO subsample ($z<1.43$). These results are presented as a robustness check against possible redshift-dependent effects in the quasar calibration. 

For completeness, we provide the full marginalized constraints for Cases A/B/C under $\Lambda$CDM, $w$CDM and CPL, as well as representative posterior contours comparing the full and low-$z$ selections.

\begin{table*}
\centering
\caption{Marginalized constraints for the $\Lambda$CDM model for the three controlled cases using the Quasars low-$z$ sample. For broad, non-Gaussian posteriors, square brackets denote central 68\% marginalized credible intervals.}
\label{tab:lcdm_cases_ABC_lowz}
\begin{tabular}{lccc}
\toprule
Parameter & Case A: QSO+SN+SH0ES & Case B: SN+SH0ES & Case C: QSO+SN (no SH0ES) \\
\midrule
$\Omega_m$   & $0.336^{+0.018}_{-0.018}$ & $0.334^{+0.018}_{-0.018}$ & $0.336^{+0.018}_{-0.018}$ \\
$H_0$        & $73.027^{+1.034}_{-1.038}$ & $73.040^{+1.037}_{-1.018}$ & $[52.283,\,91.750]$ \\
$\beta$      & $8.004^{+0.476}_{-0.476}$  & ---                        & $8.020^{+0.483}_{-0.480}$ \\
$\gamma$     & $0.593^{+0.016}_{-0.016}$  & ---                        & $0.593^{+0.016}_{-0.016}$ \\
$\delta$     & $0.238^{+0.006}_{-0.005}$  & ---                        & $0.238^{+0.006}_{-0.005}$ \\
$M_{\rm SN}$ & $-19.258^{+0.031}_{-0.032}$& $-19.258^{+0.031}_{-0.031}$& $[-19.983,\,-18.763]$ \\
\bottomrule
\end{tabular}
\end{table*}

\begin{table*}
\centering
\caption{Marginalized constraints for the $w$CDM model for the three controlled cases using the Quasars low-z ($z\leq 1.43$) sample. For broad, non-Gaussian posteriors, square brackets denote central 68\% marginalized credible intervals.}
\label{tab:wcdm_cases_ABC_lowz}
\begin{tabular}{lccc}
\toprule
Parameter & Case A: QSO+SN+SH0ES & Case B: SN+SH0ES  & Case C: QSO+SN (no SH0ES) \\
\midrule
$\Omega_m$        & $0.310^{+0.061}_{-0.072}$ & $0.291^{+0.064}_{-0.076}$ & $0.308^{+0.061}_{-0.072}$ \\
$w$               & $-0.934^{+0.145}_{-0.164}$ & $-0.893^{+0.140}_{-0.157}$ & $-0.925^{+0.141}_{-0.162}$ \\
$H_0$             & $73.024^{+1.053}_{-1.018}$ & $73.054^{+1.035}_{-1.052}$ & $[43.087,\,88.853]$ \\
$\beta$           & $7.988^{+0.481}_{-0.480}$  & ---                        & $8.039^{+0.497}_{-0.492}$ \\
$\gamma$          & $0.593^{+0.017}_{-0.017}$  & ---                        & $0.594^{+0.016}_{-0.017}$ \\
$\delta$          & $0.238^{+0.006}_{-0.006}$  & ---                        & $0.238^{+0.006}_{-0.006}$ \\
$M_{\rm SN}$      & $-19.255^{+0.032}_{-0.032}$& $-19.253^{+0.032}_{-0.033}$& $[-20.401,\,-18.830]$ \\
\bottomrule
\end{tabular}
\end{table*}

\begin{table*}
\centering
\caption{Marginalized constraints for the CPL model for the three controlled cases using the Quasars low-$z$ sample. For broad, non-Gaussian, or prior-affected posteriors, such as $w_a$ and the Case C posteriors of $H_0$ and $M_{\rm SN}$, square brackets denote central 68\% marginalized credible intervals.}
\label{tab:cpl_cases_ABC_lowz}
\begin{tabular}{lccc}
\toprule
Parameter & Case A: QSO+SN+SH0ES & Case B: SN+SH0ES & Case C: QSO+SN (no SH0ES) \\
\midrule
$\Omega_m$   & $0.400^{+0.072}_{-0.136}$ & $0.372^{+0.086}_{-0.150}$ & $0.401^{+0.073}_{-0.133}$ \\
$w_0$        & $-0.908^{+0.154}_{-0.162}$ & $-0.884^{+0.144}_{-0.158}$ & $-0.907^{+0.150}_{-0.166}$ \\
$w_a$        & $[-4.90,\,0.40]$           & $[-4.21,\,0.49]$           & $[-5.02,\,0.36]$ \\
$H_0$        & $73.051^{+1.028}_{-1.045}$ & $73.066^{+1.025}_{-1.031}$ & $[49.94,\,90.40]$ \\
$\beta$      & $8.026^{+0.477}_{-0.478}$  & ---                        & $8.058^{+0.499}_{-0.492}$ \\
$\gamma$     & $0.592^{+0.016}_{-0.016}$  & ---                        & $0.592^{+0.017}_{-0.017}$ \\
$\delta$     & $0.238^{+0.006}_{-0.005}$  & ---                        & $0.238^{+0.006}_{-0.005}$ \\
$M_{\rm SN}$ & $-19.250^{+0.032}_{-0.033}$& $-19.250^{+0.032}_{-0.033}$& $[-20.07,\,-18.79]$ \\
\bottomrule
\end{tabular}
\end{table*}

The results are broadly consistent with the trends discussed in Section~5. 
As expected, the low-$z$ selection leads to larger uncertainties due to reduced statistics, while the overall behavior of the joint calibration--cosmology inference remains stable.

\begin{figure*}
   \centering
   \includegraphics[scale=0.46]{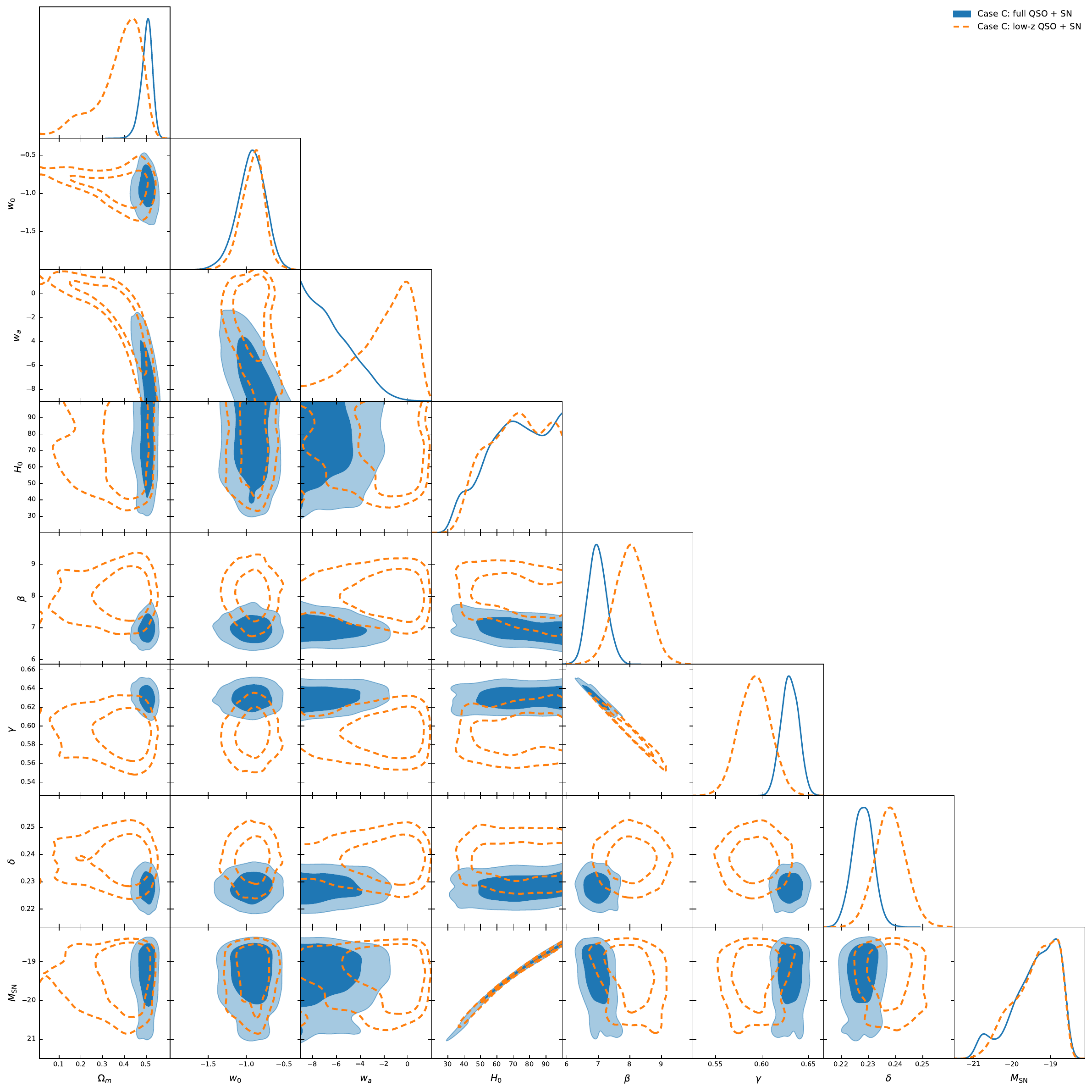}
    \caption{Comparison of CPL posterior constraints obtained from the full QSO sample and the low-$z$ subsample ($z<1.43$) within the joint QSO+SN framework (Case C). Filled contours correspond to the full sample, while dashed contours show the low-$z$ selection. Contours indicate the 68\% and 95\% credible regions.
}
   \label{ref:Full QSO-SN-cpl }
\end{figure*}

Figure~\ref{ref:Full QSO-SN-cpl } compares the CPL posterior contours obtained from the full QSO sample and the low-$z$ subsample within the same joint framework. As expected, restricting the analysis to the low-$z$ subsample weakens the constraining power and leads to broader posteriors, with shifts in the preferred regions that are most apparent in the CPL dark-energy sector. This behavior is consistent with the sensitivity to redshift coverage discussed in Section~5.4.

%%%%%%%%%figure%%%%%%%%%%%%%%%
%%%%%%%%%%%%%%%%%%%%%%%%%%%%%%%

Overall, these additional results confirm that our main conclusions, namely, the stability of the joint calibration--cosmology inference and the sensitivity to QSO redshift coverage, remain valid when restricting the analysis to the low-$z$ subsample.

% Don't change these lines
\bsp	% typesetting comment
\label{lastpage}
\end{document}